\documentclass[12pt ]{article}
\usepackage{jheppub_kim}
  \usepackage [ ]{inputenc}
 \usepackage{subcaption}
 \usepackage{adjustbox}
\usepackage{amsmath}
\usepackage{amssymb}
\usepackage{dcolumn}
\usepackage{bm}
\usepackage{color}
\usepackage{epsfig}
\usepackage{amsfonts}
\usepackage{graphicx} 
 \usepackage{epstopdf}
   \graphicspath{{plots/}} 
\usepackage{float}
\usepackage{mathtools} 

\begin{document} 
\title{ Perturbative and Non-Perturbative Contributions to Black Hole Thermodynamics with String Clouds and Dark Matter Backgrounds}
\author[a]{Kumar Sambhav Upadhyay,}
 \author[b,c]{Sudhaker Upadhyay\footnote{Visiting Associate, Inter-University Centre for Astronomy and Astrophysics (IUCAA) Pune-411007, Maharashtra, India}} 
 \author[a]{Bhabani Prasad Mandal}
\affiliation[a]{Department of Physics,
Banaras Hindu University,  Varanasi-221005, India}
\affiliation[b]{Department of Physics, K.L.S. College, Nawada, Magadh University, Bodh Gaya,   Bihar 805110, India}
\affiliation[c]{School of Physics, Damghan University, P.O. Box 3671641167, Damghan,  Iran}

\emailAdd{kumarsambhav121@gmail.com}
\emailAdd{sudhakerupadhyay@gmail.com; sudhaker@associates.iucaa.in} 
\emailAdd{ bhabani.mandal@gmail.com; bhabani@bhu.ac.in} 

\abstract{ 
We investigate the effects of perturbative and non-perturbative quantum corrections on the thermodynamics of black holes immersed in a perfect fluid dark matter (PFDM) background with a cloud of strings (CoS) in asymptotically anti-de Sitter spacetime. Starting from the Bekenstein– Hawking entropy as the semiclassical baseline, we incorporate two distinct classes of corrections arising from small thermal fluctuations about thermodynamic equilibrium. In the perturbative sector, we derive the logarithmically corrected entropy and systematically compute the resulting modifications to the mass, Helmholtz free energy, Gibbs free energy, heat capacity, and pressure. The stability structure of the system is analyzed through the sign behavior of the heat capacity, which reveals a transition from a thermodynamically unstable to a stable phase. In the non-perturbative sector, we introduce exponential corrections to the entropy and carry out a parallel analysis of all thermodynamic quantities. We demonstrate that non-perturbative effects are negligible for large black holes but become significant as the horizon radius shrinks toward the Planck regime. In both sectors, we investigate the equation of state and search for a van der Waals-like critical point by examining the simultaneous vanishing of the first and second pressure derivatives with respect to thermodynamic volume; no such inflection point is found within the physically admissible domain. Our results illuminate the contrasting roles of logarithmic and exponential entropy corrections in governing the thermodynamic stability and phase structure of PFDM black holes with a CoS.}

	\maketitle
\section{Introduction}
Black holes, originally conceived as purely gravitational objects within 
the framework of general relativity have emerged over the past five 
decades as remarkably rich thermodynamic systems whose study lies at the 
intersection of general relativity, quantum field theory, and 
statistical mechanics. In classical general relativity, a black hole is 
defined as a region of spacetime from which no signal, not even light, 
can escape to future null infinity \cite{1,2}. The boundary of this 
The region is called ``event horizon", a globally defined, one-way causal 
membrane that allows matter and signals to enter but not escape. 
According to the purely classical theory, a black hole is an 
extraordinarily simple object, characterized entirely by virtue of the 
celebrated ``no-hair theorem" by just three parameters: its mass $M$, 
angular momentum $J$, and electric charge $Q$ \cite{3,4,5,6}. All other 
things, i.e. the details of the collapsing matter, its composition, its 
quantum state  is irreversibly lost to an external observer once it 
crosses the event horizon. Therefore the classical black holes, in this 
picture, are perfectly cold, perfectly absorbing objects that emit 
nothing, have no temperature, and carry no entropy.  
 
The pioneering works of Bekenstein and Hawking in the early 1970s 
established that black holes are not merely gravitational sinks but are 
genuine thermodynamic objects endowed with well-defined macroscopic 
thermodynamic quantities \cite{7,8}. The formal framework was 
crystallized by Bardeen, Carter, and Hawking who derived the four laws 
of black hole mechanics in precise mathematical analogy with the 
classical laws of thermodynamics \cite{9}. Bekenstein's seminal insight that black hole entropy is proportional to the area of the event horizon  the celebrated Bekenstein-Hawking entropy relation $S = \frac{\text{A}}{4}$ (in natural units) was a revolutionary departure from classical intuition, suggesting that the information content of a black hole is encoded on its two-dimensional boundary rather than its three-dimensional volume \cite{8}. Following this, Hawking performed the first rigorous calculation of applying quantum field theory in curved space-time and predicts that black holes are not entirely black and they emit thermal radiation at a characteristic temperature of $T_{\mathrm{H}} = \frac{\kappa}{2 \pi}$, where $\kappa$ is the surface gravity of the event horizon \cite{10}. Together with the identification of mass with internal energy and the analogy between surface gravity and temperature, these results led to the formulation of the four laws of black hole thermodynamics, in direct analogy with the classical laws of thermodynamics. These laws, particularly the first law $dM = T_{\mathrm{H}}dS + \Omega dJ + \Phi dQ$, provide a powerful thermodynamic framework for probing the structure of black holes and 
their interactions with surrounding matter and fields. Within this framework, a wide variety of black hole solutions have been discovered 
and systematically explored in the literature 
\cite{11,12,13,14,15,16,17,18,19,20,21,22,23,24}.

Despite these remarkable achievements, the Bekenstein-Hawking entropy formula $S_{0} = \frac{\text{A}}{4}$ is understood to be the leading-order, semiclassical result, valid in the limit of large black holes where quantum gravitational corrections are negligible. As a black hole evaporates via Hawking radiation, its horizon area shrinks and the temperature rises, eventually reaching a regime where quantum corrections to the entropy become significant and can no longer be ignored. Understanding these corrections is crucial both for probing the quantum structure of spacetime near the Planck scale and for addressing fundamental questions such as the black hole information paradox. Two broad classes of corrections have been studied extensively in the literature: perturbative corrections, which arise as a power series expansion in the inverse of the equilibrium entropy and include the well-known logarithmic correction terms, and non-perturbative corrections, which are exponentially suppressed in the entropy and capture effects that are invisible to any order in perturbation theory.

As a black hole loses mass through Hawking radiation, its size gradually diminishes, and the role of thermal fluctuations becomes increasingly significant. In this regime, the standard Bekenstein-Hawking entropy receives corrections, with the leading-order contribution typically taking a logarithmic form \cite{25}. Such logarithmic (perturbative) corrections induce nontrivial modifications to the thermodynamic behaviour and stability properties of the black hole system. Accordingly, significant attention has been devoted to incorporating such corrections within the framework of black hole thermodynamics, with their applicability extended to a wide spectrum of gravitational solutions, encompassing higher-dimensional and nontrivial horizon topologies such as black rings as well as diverse black hole models, including quasitopological black holes \cite{26}, van der Waals black holes in higher dimensional AdS space \cite{27}, AdS black holes in massive gravity \cite{28}, rotating and charged BTZ black holes~\cite{29}, black holes governed by $f(R)$ gravity \cite{30}, 4D AdS Einstein-Gauss-Bonnet black holes \cite{31}, charged AdS black holes \cite{32}, black branes \cite{33}, Schwarzschild–Beltrami-de Sitter black hole \cite{34}, self-gravitating Skyrmion black holes \cite{35}, Bardeen regular black holes \cite{36} Horava–Lifshitz gravity black holes \cite{37} and phantom BTZ black holes \cite{38}. Recent studies suggest that these correction terms exhibit a universal character and can be expressed, to a good approximation, in the following general form \cite{39}:
\begin{align*}
 S = S_{0} + \alpha \ln S_{0} + \frac{\gamma}{S_{0}} + \eta e^{-S_{0}} + ...   
 \end{align*}
 where dots represent the higher order corrections to unperturbed entropy. Here, we observe that at leading order, the entropy acquires a logarithmic correction, consistent with results obtained from microstate counting in both string theory and loop quantum gravity. In contrast, the exponential correction term emerges when the microstate counting is restricted to quantum states localized on the horizon.

However, by reducing more black hole size the non-perturbative corrections in entropy becomes dominant. An exponential contribution gives rise to non-perturbative corrections to the black hole entropy \cite{40}.
\begin{align*}
  S_{\mathrm{exp}} = S_{0} + \eta e^{-S_{0}},  
\end{align*} 
here, $S_{0}$ is the unperturbed black hole entropy proportional to the event horizon radius and $\eta$ is the correction factor. This represents a universal feature that can be effectively employed to probe a wide range of black hole thermodynamic properties. The effect of exponential correction is negligible for large event horizon radius but significant for infinitesimally small horizon radius. The effects of non-perturbative exponential corrections have been studied in various black hole models such as 4D Reissner-Nordström black hole \cite{41}; 3D black holes conformally coupled to scalar field \cite{42}; static dirty black holes \cite{43}; quasi-topological black holes \cite{44}; quantum sized AdS black holes \cite{45},   etc.

In this paper we first review the black holes with a CoS and PFDM. We first analyze the effects of small thermal fluctuations as a statistical perturbations around the equilibrium of PFDM black holes with CoS arising the correction terms in the original unperturbed entropy. The correction terms that arises in unperturbed entropy are of two types i.e. perturbative (logarithmic) corrections and non-perturbative (exponential) corrections. We first study the effect of perturbative (logarithmic) correction to the entropy of PFDM black holes with a CoS after that we derive the expression for corrected mass using the first law of thermodynamics and analyze its variation with respect to event horizon radius of the black hole. Following this, we calculate the first order perturbative corrected thermodynamic potentials such as Helmholtz free energy, Gibbs free energy, heat capacity and analyze their behavior as a function of the event horizon radius with and without logarithmic correction after that we also study the stability of our black hole system by analyzing the positive and negative behavior of heat capacity curves plotted against the event horizon radius. subsequently, we also derive the expression for logarithmic corrected pressure and study its behavior and variations with respect to horizon radius. For next section we do the same calculations of various thermodynamic quantities as the above section; stability analysis of heat capacity and pressure variations with respect to event horizon radius by using non-perturbative (exponential) correction introduce in original entropy.

The paper is presented in the following manner. In Sec. \ref{sec2}, we provide a brief review and discussion of the foundational aspects of the black holes with a CoS and PFDM solution.  In section \ref{sec3},we examine small statistical thermal fluctuations around equilibrium thermodynamics of PFDM black holes with a CoS which give rise to corrections to the  entropy of the black hole. Afterwards we consider perturbative or logarithmic corrected entropy then further analyze perturbative corrected behavior of all other thermodynamic quantities and heat capacity of PFDM black holes with a CoS. In subsection \ref{sec4}, we analyze the non-perturbative (exponential) calculations of entropy and other thermodynamic potentntials as well as heat capacity and its stability implications. After that, we calculate and analyze the exponentially corrrected pressure behavior as a function of the horizon radius and then search for the point of inflection of the PFDM black holes with a CoS.

\section{Black holes with a CoS and PFDM backgrounds}\label{sec2}
In this section, we revisit the black holes with a CoS and PFDM.
The Einstein-Hilbert action describing Einstein's gravity coupled to a  CoS  parameter, immersed in a  PFDM  background in an asymptotically AdS spacetime is expressed as \cite{46}
\begin{equation}
 S_{\mathrm{EH}} = \int d^4x \, \sqrt{-g} \left[R - 2\Lambda + 2 \nabla_\mu \Phi \, \nabla^\mu \Phi -4 V(\Phi) -4 \mathcal{L}_{\mathrm{DM}}  - F_{\mu\nu} F^{\mu\nu}\right] + S_{\mathrm{CS}}, 
 \label{S_pfdm}
\end{equation}
where $R$ is the scalar curvature, $\Lambda$ is the cosmological constant, $\Phi$ represents the phantom field with associated phantom field potential $V(\Phi)$, the term $\mathcal{L}_{\mathrm{DM}}$ represents the Lagrangian density of the dark matter field, $ F_{\mu\nu} = \partial A_\mu - \partial A_\nu$ is the electromagnetic field tensor, and $S_{\mathrm{CS}}$ corresponds to the action of  CoS  source.   

The  CoS  source is described by the following action:
\begin{equation}
S_{\mathrm{CS}} = \int_{\Sigma} m (-h)^{-1/2} \, d\gamma^0 \, d\gamma^1 = \int_{\Sigma} m\left(-\frac{1}{2}\, \Sigma^{\mu\nu}\Sigma_{\mu\nu}\right)^{1/2} \, d\lambda^0 \, d\lambda^1,    
\end{equation}
where $m$ is the mass of the string, $h$ is the determinant of the reduced metric, and $\gamma^0$ and $\gamma^1$ correspond to the time-like and space-like coordinates respectively. The antisymmetric tensor $\Sigma^{\mu\nu}$ is defined as
\begin{equation}
 \Sigma^{\mu\nu} = \epsilon^{ab} \, \frac{\partial y^{\mu}}{\partial y^{a}} \, \frac{\partial y^{\nu}}{\partial y^{b}},   
\end{equation}
where $\epsilon^{ab}$ denotes the Levi-Civita tensor with the non-vanishing component $\epsilon^{01} = -\epsilon^{-10} = 1$.

Varying the action (\ref{S_pfdm}) with respect to metric tensor $g_{\mu\nu}$ and electromagnetic potential $A_\mu$ yield the following equations of motion:
\begin{equation}
G_{\mu\nu} + \Lambda g_{\mu\nu} = T^{\mathrm{CS}}_{\mu\nu} + T_{\mu\nu},
\label{G_munu}
\end{equation}

\begin{equation}
 \nabla_{\mu} F^{\mu\nu} = 0 \quad \text{and} \quad \nabla_{\nu} (*F^{\mu\nu}) = 0,
\end{equation}
where $A_{t} = -\frac{Q}{r}$ is the non-vanishing component of the electromagnetic potential $A_\mathrm{\mu}$. The energy momentum tensor ($EMT$) for the CoS   source can be expressed as
\begin{equation}
 T_{\mathrm{CS}}^{\mu\nu} = \frac{\rho \, \Sigma^{\mu\rho} \Sigma^{\,\nu}_{\rho}}{\sqrt{-h}} = \frac{a}{r^2} \, \operatorname{diag}[1,1,0,0],
\end{equation}
where $a$ is the  CoS  parameter.

Next, we express the $EMT$ corresponding to the PFDM in the presence of electromagnetic fields as
\begin{equation}
 T_{\mathrm{\mu\nu}} = 2 \left[F_{\mathrm{\mu\gamma}}F_{\nu}^{\gamma} - \frac{1}{2} g_{\mathrm{\mu\nu}}\right] + 2\nabla_{\mu} \Phi \nabla_{\nu} \Phi - g_{\mathrm{\mu\nu}} \nabla_{\gamma} \Phi \nabla^\gamma \Phi + T_{\mathrm{\mu\nu}}^{\mathrm{DM}} ,
 \label{T_munu}
\end{equation}
where $T_{\mathrm{\mu\nu}}^{\mathrm{DM}}$ represents the $EMT$ of the PFDM field.

In order to obtain the black hole solution, we consider the most general static, spherically symmetric spacetime metric, which can be written as
\begin{equation}
 ds^2 = -e^A(r)dt^2 + e^B(r)dr^2 + r^2 \left(d\theta^2 + \sin^2{\theta}d\phi^2\right).  
\end{equation}
For the static case, the Einsteins equation read
\begin{align}
G^{t}_{ t} &= e^{-B(r)} \left( \frac{1}{r^2} - \frac{B'(r)}{r} \right) - \frac{1}{r^2} - \Lambda r^2 = T^{t}_{t}, \label{G_tt} \\
G^{r}_{ r} &= e^{-B(r)} \left( \frac{1}{r^2} - \frac{A'(r)}{r} \right) - \frac{1}{r^2} - \Lambda r^2 = T^{r}_{ r}, \label{G_rr} \\
G^{\theta}_{ \theta} &= \frac{e^{-B(r)}}{2} \left( A''(r) + \frac{A'(r)^2}{2} + \frac{A'(r) - B'(r)}{r} - \frac{A'(r) B'(r)}{2} \right) - \Lambda r^2 = T^{\theta}_{ \theta}, \label{G_theta}
\end{align}
where, prime denotes the differentiation with respect to $r$. Here, we observe that the components, $G^{\theta}_{\theta}$ and 
$G^{\phi}_{\phi}$, of the Einstein field equation (\ref{G_munu}) are same.

The components of $EMT$ (\ref{T_munu}) are
\begin{align}
T^{t}_{ t} &= \frac{Q^2}{r^4} + \frac{a}{r^2} + \frac{1}{2} e^{-B(r)} \Phi'^2 - V(\Phi) - \rho_{\mathrm{DM}}, \\
T^{r}_{ r} &= \frac{Q^2}{r^4} + \frac{a}{r^2} - \frac{1}{2} e^{-B(r)} \Phi'^2 - V(\Phi), \\
T^{\theta}_{ \theta} &= \frac{1}{2} e^{-B(r)} \Phi'^2 - V(\Phi).
\end{align}
In order to find the black hole solution that follows the condition $A(r) = -B(r)$ ($g_{{tt}} = -\frac{1}{g_{{rr}}}$) and $\rho_{\mathrm{DM}} = e^{A(r)} \Phi'^2 > 0$, we set $B(r) = \ln\left(1 - \mathcal{U}(r)\right)$ and substitute this into Eqs.~(\ref{G_tt}) and (\ref{G_theta}). We, then, obtain
\begin{equation}
 r^2\mathcal{U''}(r) + 2\epsilon r \mathcal{U'}(r) + 2\left(1 - \epsilon\right) \mathcal{U}(r) - \Lambda r^2 = \frac{a}{r^2} + \frac{Q^2}{r^4},
 \label{r_U}  
\end{equation}
where we have set $T^{\theta\theta} + T^{\theta\theta}_{\mathrm{CS}} = T^{\phi\phi} + T^{\phi\phi}_{\mathrm{CS}} = \left(1 - \epsilon\right) T^{tt} + T^{tt}_{\mathrm{CS}}$, with a constant $\epsilon$. Now, the equation (\ref{r_U}) leads to the following solutions:
\begin{align}
 \mathcal{U}(r) = \frac{2M}{r} - \frac{r^2(1-\epsilon)}{r} - \frac{Q^2}{r^2} + a, \qquad \epsilon \neq \frac{3}{2} \\
 \mathcal{U}(r) = \frac{2M}{r} + \frac{\lambda}{r} \ln\left(\frac{r}{\lambda}\right) - \frac{Q^2}{r^2} + a, \qquad \epsilon = \frac{3}{2},
\end{align}
where $\lambda$ is the integration constant. The black hole solution for $\epsilon = \frac{3}{2}$ is given by \cite{46}
\begin{align}
ds^2 &= -\left(1 - \frac{2M}{r} + \frac{Q^2}{r^2} - a 
+ \frac{\lambda}{r} \ln\left(\frac{r}{\lambda}\right) 
+ \frac{r^2}{l^2}\right) dt^2 \nonumber \\
&\quad + \left(1 - \frac{2M}{r} + \frac{Q^2}{r^2} - a 
+ \frac{\lambda}{r} \ln\left(\frac{r}{\lambda}\right) 
+ \frac{r^2}{l^2}\right)^{-1} dr^2
 + r^2 d\Omega^2.
 \label{ds_pfdmcos}
\end{align}
The black hole solution presented in (\ref{ds_pfdmcos})  obtained in the presence of an electromagnetic field, the  CoS  parameter, and the  PFDM  field contribution, arises as a consistent solution of the Einstein field equations from (\ref{G_tt}) to (\ref{G_theta}). This solution is specified by five independent parameters, namely the mass $M$, electric charge $Q$,  CoS parameter $a$, scale parameter $\lambda$, and the cosmological constant $\Lambda$. The latter is related to the AdS  length scale $l$ through the relation $\Lambda = -3/l^2$.

The obtained black hole solution generalizes several known configurations and exhibits the correct limiting behavior in appropriate parameter regimes. In particular, in the absence of the scale parameter $\lambda$, the solution smoothly reduces to the charged Letelier black hole \cite{47,48,49}. Furthermore, in the limit $Q = 0$, the electromagnetic contribution vanishes, and the solution reduces to
\begin{align}
ds^2 &= -\left(1 - \frac{2M}{r}  - a 
+ \frac{\lambda}{r} \ln\left(\frac{r}{\lambda}\right) 
+ \frac{r^2}{l^2}\right) dt^2 \nonumber \\
&\quad + \left(1 - \frac{2M}{r}  - a 
+ \frac{\lambda}{r} \ln\left(\frac{r}{\lambda}\right) 
+ \frac{r^2}{l^2}\right)^{-1} dr^2
 + r^2 d\Omega^2.
 \label{ds_pfdmrn}
\end{align}
The black hole solution (\ref{ds_pfdmcos}) exhibits the appropriate limiting behaviours. In particular, it reduces to the Reissner-Nordstr\"om black hole for $\lambda = a =0$, to the Letilier black hole in the limit $\lambda =0, Q = 0$, and to the Schwarzschild black hole when $\lambda = a = Q =0$. Furthermore, the inclusion of a negative cosmological constant ensures that the spacetime is asymptotically AdS. 

Now, the entropy associated with the black hole solution (\ref{ds_pfdmcos}) is obtained from the first law of thermodynamics as
\begin{equation}
 dM = T_{{H}} dS_{0} + \phi dQ + \lambda \Lambda_{1} + PdV , 
\end{equation}
where $\Lambda_{1}$ is conjugate to $\lambda$.

For constant charge ($Q$) the entropy of our given black hole system can be calculated as 
\begin{equation}
 S_{0} = \int \frac{1}{T_{{H}}} \frac{dM}{dr_{+}} dr_{+} = \pi r_{+}^2 = \frac{A}{4}.
 \label{S_pfdmcos}
\end{equation}
The entropy follows Bekenstein-Hawking area law, with the associated Smarr relation expressed as
\begin{equation}
 M = 2 T_{{H}} S_{0} - 2PV + \phi Q + \lambda \Lambda_{1},  \end{equation} 
with terms having following explicit analytical expressions:
\begin{equation}
\phi = \frac{dM}{dQ} = \frac{Q}{r_{+}}, \quad V = \frac{4}{3} \pi r^3_{+},    
\end{equation}
\begin{equation}
 \Lambda_{1} = -\frac{1}{2} + \frac{1}{2} \ln\left[\frac{r_{+}}{\lambda}\right].   
\end{equation}

\section{Perturbative corrections to thermodynamics of PFDM black holes with a CoS}\label{sec3}
The perturbative logarithmic corrected entropy of black holes in general, is given by \cite{25}:
\begin{equation}
 S_{c} = S_{0} - \alpha \ln(S_{0} T_H^2),
 \label{S_lnentr}  
\end{equation}
where $S_{0}$ is the unperturbed entropy of any black hole system, $T_{H}$ is the Hawking temperature of the black hole and $\alpha$ denotes the correction parameter associated with small statistical thermal fluctuations around the equilibrium of black hole. This parameter takes the value $\alpha = \frac{1}{2}$ when thermal fluctuations are taken into account and vanishes that is $\alpha = 0$ in the absence of such fluctuations.

Now, the unperturbed entropy $S_{0}$ of the PFDM black holes surrounded by a CoS can be written by using Eq. (\ref{S_pfdmcos}) as \cite{46} 
\begin{equation}
 S_{0} = \frac{A}{4} = \pi r^2_{+}.  
\end{equation}
Also, by using metric  function $f(r)$ given by Eq. (\ref{ds_pfdmcos}), and employing the standard relation $T_{\mathrm{H}} = \frac{f'(r)}{4\pi}$ the Hawking temperature $T_{H}$ of PFDM black holes with a CoS is calculated as \cite{45} 
\begin{equation}
 T_{H} = \frac{1}{4 \pi r_{+}} \Bigg(1 - a + \frac{3 r^2_{+}}{l^2} - \frac{Q^2}{r^2_{+}} + \frac{\lambda}{r_{+}}\Bigg),
 \label{T_pfdmHaw}
\end{equation}
where $a$ is the CoS   parameter, $l$ is the AdS length, $Q$ is the charge and $\lambda$ is the scale parameter. 

Therefore, from Eq. (\ref{S_lnentr}), we find that
\begin{equation}
 S_{c} = \pi r^2_{+} - \alpha \ln \Bigg[\pi r^2_{+} \Bigg( \frac{1}{4 \pi r^2_{+}} \left(1 - a + \frac{3 r^2_{+}}{l^2} - \frac{Q^2}{r^2_{+}} + \frac{\lambda}{r_{+}}\right)\Bigg)^2\Bigg].
 \label{lncoentr}
\end{equation}
 
\begin{figure}[htb]

\begin{tabular}{c c} 
    \includegraphics[width=0.52\textwidth]{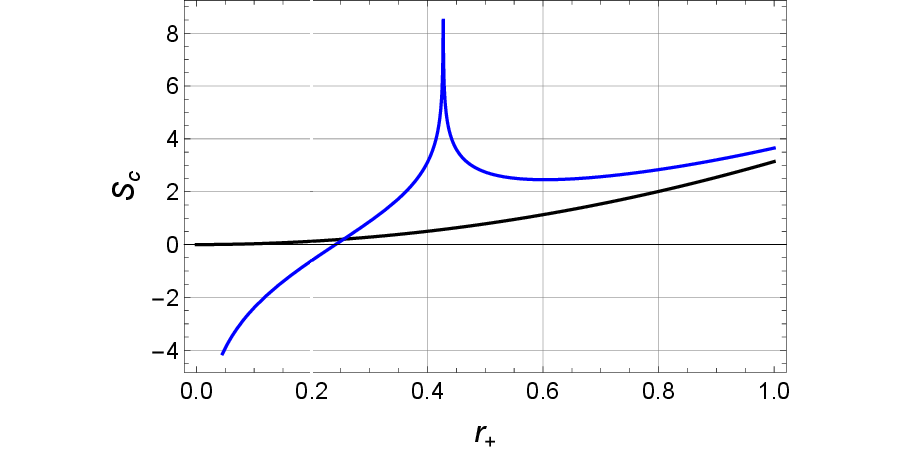}
    \includegraphics[width=0.52\textwidth]{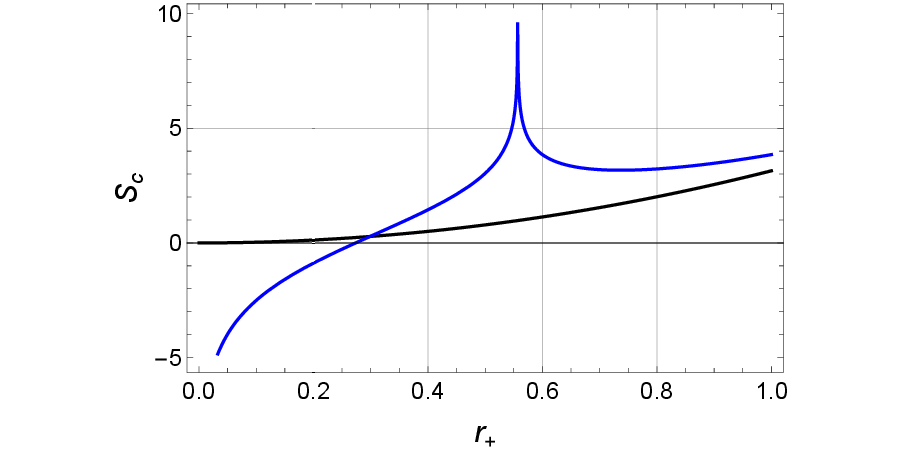}
    \end{tabular} 
    \caption{Perturbative corrected entropy vs event horizon radius of black holes with a CoS and PFDM. Here, $\alpha = 0$ is represented by black curve and $\alpha = 0.50$ is represented by blue curve. Left panel: $a = 0.75$ and $\lambda = 2$. Right panel: $a = 0.50$ and $\lambda = 1$. Both cases are plotted for constant values of $Q = l =1$.}
    \label{fig1}
\end{figure}
 
The behavior of the perturbatively corrected entropy  as a function of the event horizon radius  is illustrated in   Fig. \ref{fig1}. Here, we find that the variation of the perturbatively corrected entropy \(S_{c}\) with the event horizon radius \(r_{+}\) for black holes surrounded by a CoS and PFDM, where the black curve corresponds to the uncorrected case \((\alpha =0)\) and the blue curve represents the corrected case \((\alpha =0.50)\). In the left panel \((a=0.75,\ \lambda =2)\), the uncorrected entropy increases monotonically with \(r_{+}\), while the corrected entropy becomes negative for small horizon radius and exhibits a sharp divergence around \(r_{+}\approx 0.43\), after which it decreases abruptly and gradually stabilizes for larger \(r_{+}\). A similar behavior is observed in the right panel for \((a=0.50,\ \lambda =1)\), where the divergence shifts to a comparatively larger horizon radius \(r_{+}\approx 0.50\). The results indicate that perturbative corrections significantly modify the thermodynamic behavior in the small black hole regime and may lead to critical behavior or phase transitions, whereas the uncorrected entropy remains smooth and regular throughout. We plot both panels  for fixed values of \(Q=l=1\).

 {The logarithmically corrected entropy of the PFDM black hole surrounded by a cloud of strings exhibits a sharp but finite peak at a particular horizon radius $r_{+}$.This feature possesses important thermodynamic and physical significance, since it marks the vicinity of the extremal configuration of the black hole and indicates the onset of critical thermodynamic behavior. Since the logarithmic correction depends explicitly on $T^2_{\mathrm{H}}$, the entropy becomes highly sensitive to the regime where the Hawking temperature approaches to zero.}
The finite entropy peak occurs near the horizon radius satisfying 
\begin{align*}
 1 - a + \frac{3 r^2_{+}}{l^2} - \frac{Q^2}{r^2_{+}} + \frac{\lambda}{r_{+}} \approx 0 .
\end{align*}
This corresponds to the near-extremal condition $T_{H} \approx 0$. 

 {Unlike an exact mathematical divergence, the entropy peak obtained in the present analysis remains finite. This occurs because the numerical computation approaches the extremal point very closely but does not exactly satisfy $T_{H} = 0$ Consequently, the logarithmic term becomes very large but remains finite, producing a sharp entropy enhancement rather than an actual singularity. The finite nature of the peak therefore indicates a physically accessible near-extremal regime rather than a mathematically singular thermodynamic divergence.}
 {The appearance of this sharp entropy enhancement also reflects the increasing dominance of quantum thermal fluctuations near extremality. In the vicinity of $T_{H} \to 0$ the semiclassical equilibrium description becomes highly sensitive to small perturbations, causing the logarithmic correction term to grow rapidly. Therefore the finite entropy peak can be interpreted as a signature of the near-extremal thermodynamic state and enhanced quantum fluctuations of the PFDM black holes with a CoS. }

\subsection{Variation of peak value of entropy of PFDM black holes with a CoS as a function of its parameters}
\begin{figure}[htb]
\begin{tabular}{c c} 
    \includegraphics[width=0.52\textwidth]{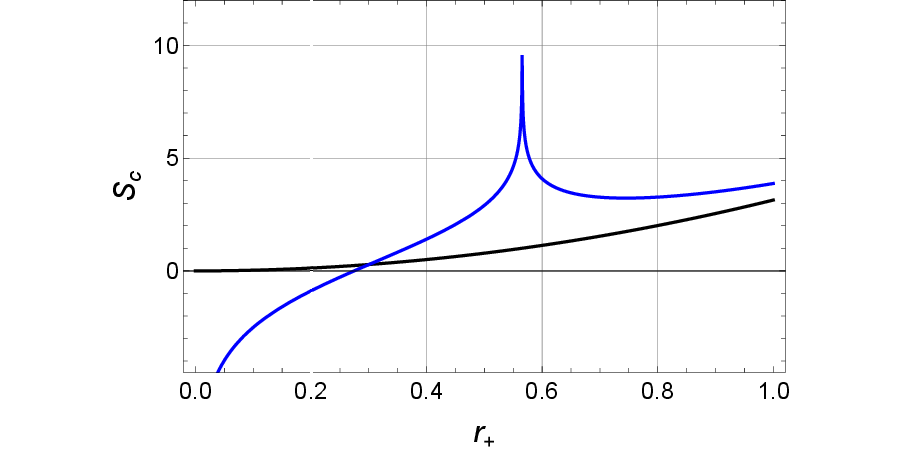}
    \includegraphics[width=0.52\textwidth]{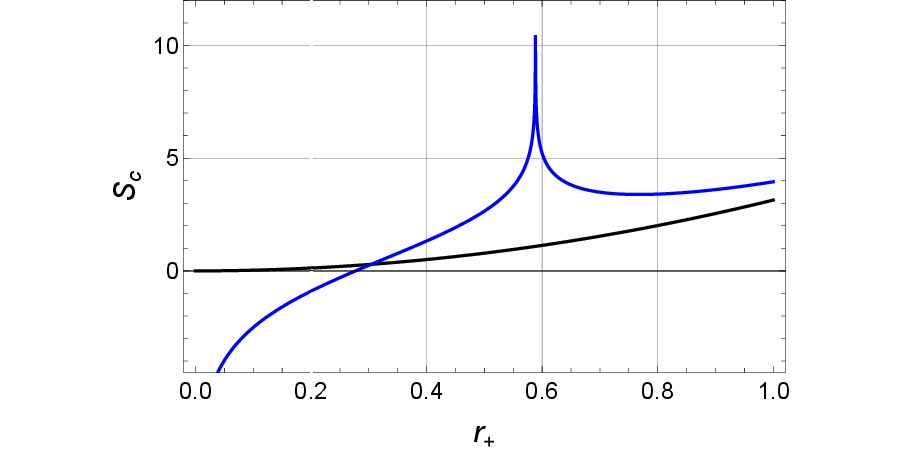}
    \end{tabular} 
    \caption{variation of Perturbative corrected entropy vs event horizon radius of black holes with a CoS parameter. Here, $\alpha = 0$ is represented by black curve and $\alpha = 0.50$ is represented by blue curve. Left panel: $a = 0.60$ . Right panel: $a = 0.85$ . Both cases are plotted for constant values of $Q = l = \lambda = 1$.}
    \label{figa1}
\end{figure}

 {Here, we describe the variation of the entropy peak of PFDM black hole with CoS as a function of its cloud of strings parameter `a' keeping all other parameters as constant. From the figure \ref{figa1} we see that entropy peak slightly shift towards rightwards i.e. in increasing value of $r_{+}$ with increased value of maximum entropy.}

\begin{figure}[htb]
\begin{tabular}{c c} 
    \includegraphics[width=0.52\textwidth]{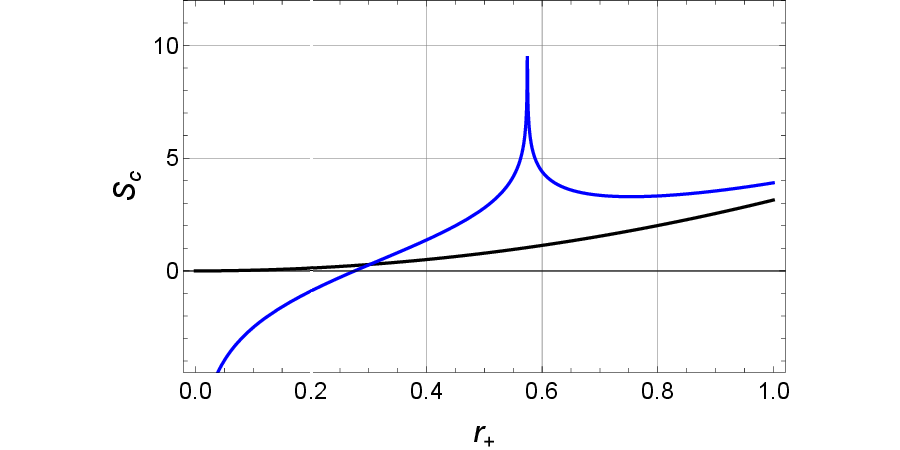}
    \includegraphics[width=0.52\textwidth]{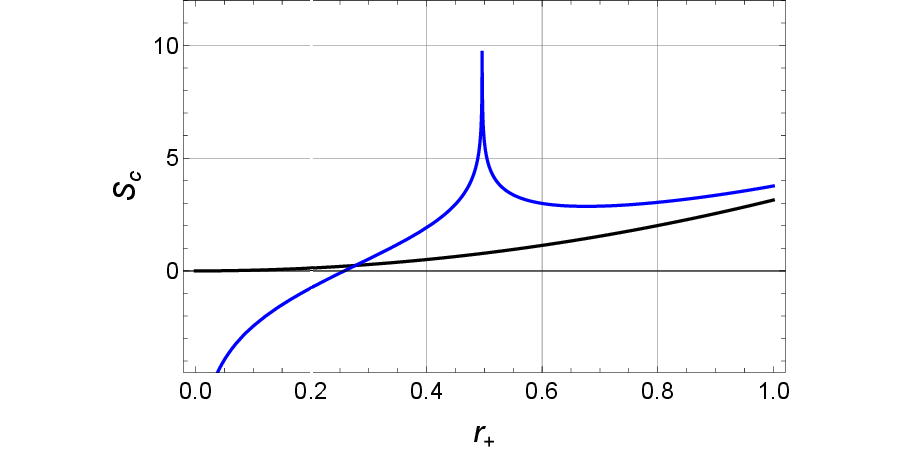}
    \end{tabular} 
    \caption{variation of Perturbative corrected entropy vs event horizon radius of black holes with a PFDM scale parameter. Here, $\alpha = 0$ is represented by black curve and $\alpha = 0.50$ is represented by blue curve. Left panel: $\lambda = 1.0$ with $a = 0.70$ . Right panel: $\lambda = 1.5$ with $a = 0.70$ . Both cases are plotted for constant values of $Q = l = 1$.}
    \label{figl1}
\end{figure}

 {From the figure \ref{figl1} we illustrate the variation of the entropy peak of PFDM black hole with CoS as a function of its scale parameter $\lambda$ treating all other parameters as constants. Here, we observe from the figure \ref{figl1} that entropy peak shifts towards leftwards in the decreasing value of $r_{+}$ with increased value of maximum entropy. }

\subsection{Perturbative corrected mass}
\begin{figure}[htb]
\centering
\begin{tabular}{cc} 
    \includegraphics[width=0.52\textwidth]{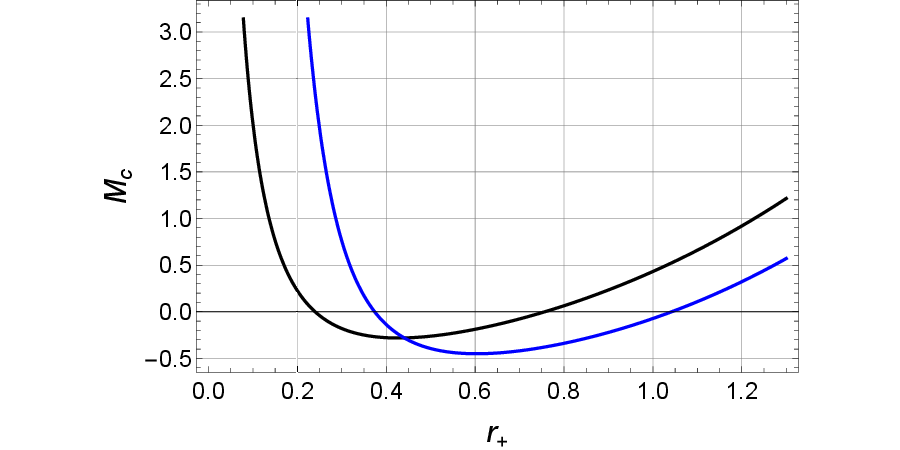}
    \includegraphics[width=0.52\textwidth]{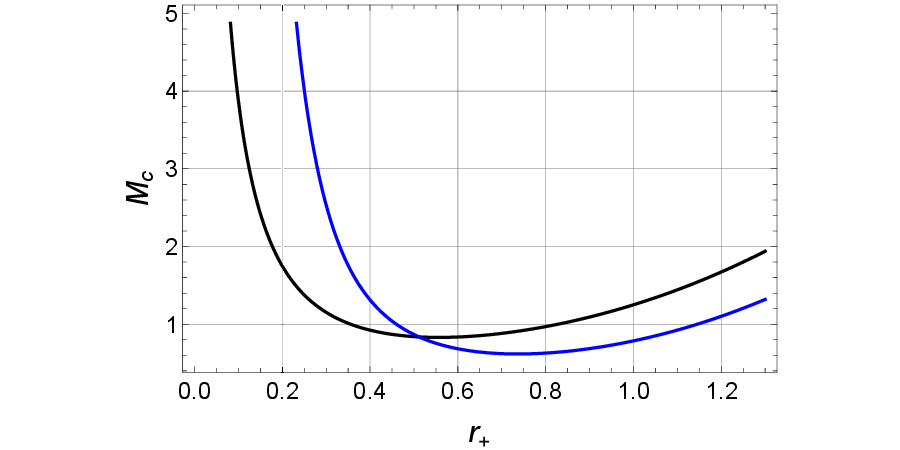}
    \end{tabular} 
    \caption{Logarithmic corrected mass vs event horizon radius of black holes with a CoS and PFDM. Here, $\alpha = 0$ is represented by black curve and $\alpha = 0.50$ is represented by blue curve. Left panel: $a = 0.75$ and $\lambda = 2$. Right panel: $a = 0.50$ and $\lambda = 1$. Both cases are plotted for constant values of $Q = l =1$.}
    \label{M_c}
\end{figure}

The first law of thermodynamics, when perturbative correction is taken into account, is expressed as 
\begin{equation}
 dM_{c} = T_{H} dS_{c} + \phi dQ + PdV,
 \label{M_lnc}
 \end{equation}
where $Q$ is the charge, $\phi$ is its conjugate quantity potential and $P$ is the pressure. Also, since charge $Q$ and volume $V = \frac{4}{3} \pi r^3_{+}$ are  constants their variations vanish i.e. $dQ = 0$ and $dV = 0$. 
Therefore, upon integration  this equation leads to corrected mass $M_{c}$  as
\begin{equation}
 M_c = \int T_{H} \, dS_c. 
\end{equation}
Substituting the values of $T_{{H}}$ from (\ref{T_pfdmHaw}) and $S_{c}$ from (\ref{lncoentr}), we get the following explicit expression: 
 \begin{eqnarray}
M_{c}=
\frac{r_{+}^{3}}{2l^{2}}
+\frac{Q^{2}+r_{+}^{2}-a r_{+}^{2}}{2r_{+}}
+\frac{\lambda}{2}\log\!\left(\frac{r_{+}}{\lambda}\right)
+\alpha\left[
-\frac{3r_{+}}{\pi l^{2}}
+\frac{4Q^{2}-3r_{+}\lambda}{12\pi r_{+}^{3}}
\right].\label{M_cpfdm}
\end{eqnarray}
We illustrates the variation of the logarithmic corrected mass \(M_{c}\) with the event horizon radius \(r_{+}\) for black holes surrounded by a CoS and PFDM in figure \ref{M_c}, where the black curve corresponds to the uncorrected case \((\alpha =0)\) and the blue curve represents the logarithmically corrected case \((\alpha =0.50)\). In the left panel \((a=0.75,\ \lambda =2)\), we find that the corrected mass initially decreases in the small horizon radius region and then increases rapidly with \(r_{+}\), eventually approaching the behavior of the uncorrected mass for larger black holes.  In fact, the uncorrected mass remains positive and increases monotonically throughout the physical domain. In the right panel \((a=0.50,\ \lambda =1)\), we observe a similar qualitative behavior, although the growth rate of the corrected mass is comparatively smoother and the deviation between corrected and uncorrected curves becomes less pronounced for intermediate values of \(r_{+}\). Here, we demonstrate that logarithmic corrections significantly affect the thermodynamic mass in the small black hole regime, while their influence gradually diminishes as the horizon radius increases.  

 {The intersection point, where the corrected and uncorrected black hole mass curves cross each other,   suggests that corrected perturbative contribution originated from statistical thermal fluctuations to the black hole mass becomes effectively zero or vanishes at that particular event horizon radius. Therefore we can say that the intersection point between the corrected and uncorrected mass curves corresponds to a particular horizon radius at which the contribution of small statistical thermal fluctuations to the black hole mass vanishes effectively. This point separates the thermal fluctuation-dominated regime of small black holes from the regime where the PFDM black hole with CoS is predominantly governed by the uncorrected thermodynamic behavior. }

\subsection{Helmholtz free energy}
The corrected Helmholtz free energy is given by 
\begin{equation}
 F_{c} = - \int S_{c} \, dT_{H}.
 \end{equation}
This leads to
\begin{eqnarray}
F_{c} &=& 
\frac{3Q^{2}}{4r_{+}}
-\frac{r_{+}^{3}}{4l^{2}}
+\frac{r_{+}}{4}(1-a)
+\frac{\lambda}{2}\ln\!\left(\frac{r_{+}}{\lambda}\right)+
\frac{\alpha Q^{2}}{3\pi r_{+}^{3}}
-\frac{3\alpha r_{+}}{\pi l^{2}}
-\frac{\alpha\lambda}{4\pi r_{+}^{2}}
\nonumber\\
&+&
\frac{\alpha}{4\pi r_{+}^{3}}
\left(
\frac{3r_{+}^{4}}{l^{2}}
-Q^{2}
+r_{+}^{2}(1-a)
+\lambda r_{+}
\right)
\ln\!\left[
\frac{1}{16\pi}
\left(
-1+a+\frac{Q^{2}}{r_{+}^{2}}
-\frac{3r_{+}^{2}}{l^{2}}
-\frac{\lambda}{r_{+}}
\right)^{2}
\right].
\label{F_cpfdm}
\end{eqnarray}
 \begin{figure}[htb]
\centering
\begin{tabular}{cc} 
    \includegraphics[width=0.52\textwidth]{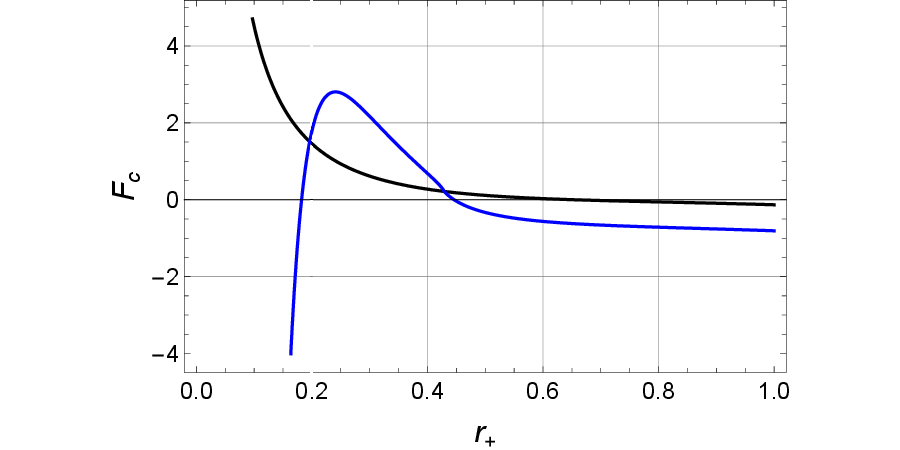}
    \includegraphics[width=0.52\textwidth]{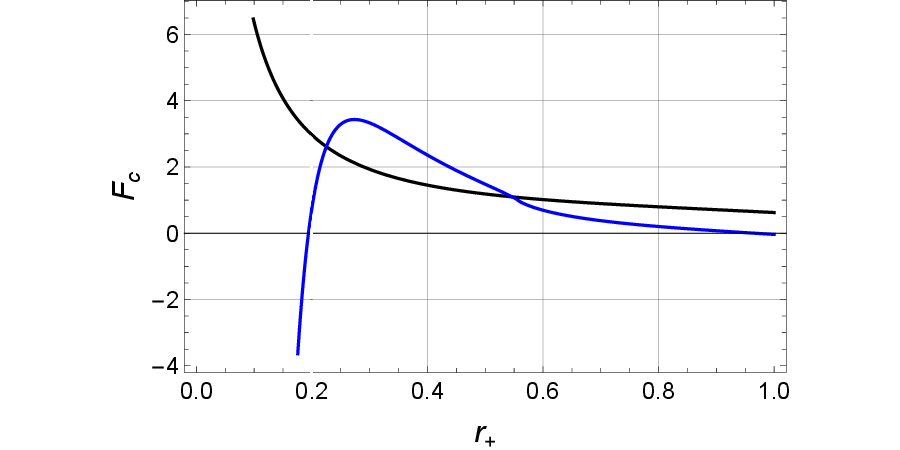}
    \end{tabular} 
    \caption{Logarithmic corrected Helmholtz free energy vs event horizon radius of black holes with a CoS and PFDM. Here, $\alpha = 0$ is represented by black curve and $\alpha = 0.50$ is represented by blue curve. Left panel: $a = 0.75$ and $\lambda = 2$. Right panel: $a = 0.50$ and $\lambda = 1$. Both cases are plotted for constant values of $Q = l =1$.}
    \label{lncohpfdm}
\end{figure}
From the Fig. \ref{lncohpfdm}, we observe that the logarithmic corrected Helmholtz free energy \(F_{c}\) exhibits significant variation with the event horizon radius \(r_{+}\) for black holes surrounded by a CoS and PFDM. The black curve represents the uncorrected case \((\alpha =0)\), while the blue curve corresponds to the logarithmically corrected case \((\alpha =0.50)\). In the left panel \((a=0.75,\ \lambda =2)\), the corrected Helmholtz free energy initially decreases in the small horizon radius region, reaches a minimum value, and then gradually increases with increasing \(r_{+}\), approaching the behavior of the uncorrected curve for larger black holes. In the right panel \((a=0.50,\ \lambda =1)\), a similar trend is observed, although the variation of the corrected free energy becomes comparatively smoother and the deviation between the corrected and uncorrected curves reduces for intermediate and large values of \(r_{+}\). These results indicate that logarithmic corrections strongly influence the thermodynamic behavior in the small black hole regime, whereas their effect becomes less significant as the event horizon radius increases.  {In the context of black hole thermodynamics Helmholtz free energy tells whether the black hole configuration is thermodynamically favored or not. Lower or negative Helmholtz free energy implies more thermodynamically stable state and higher or positive value implies less stable state/configuration. }

 {At the intersection point of corrected and uncorrected Helmholtz free energy curves we find that   contribution coming from statistical thermal fluctuations effectively becomes zero at that particular horizon radius $r_{+}$. In other words the intersection point between the corrected and uncorrected Helmholtz free energy curves represents a particular horizon radius at which the perturbative corrections originated from statistical thermal fluctuations effectively vanishes. This separates the regime dominated by quantum/statistical corrections from the regime where the classical thermodynamic description becomes comparatively dominant. }
\subsection{Gibbs free energy}
\begin{figure}[htbp]
\centering
\begin{tabular}{cc} 
    \includegraphics[width=0.52\textwidth]{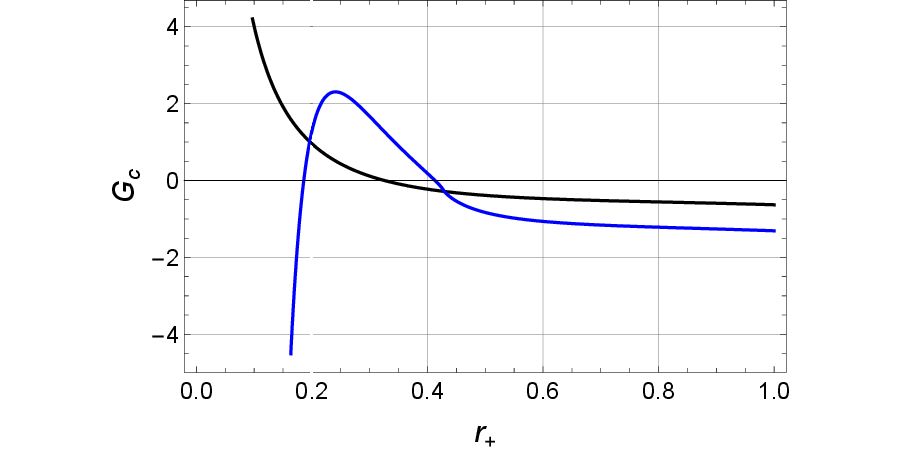}
    \includegraphics[width=0.52\textwidth]{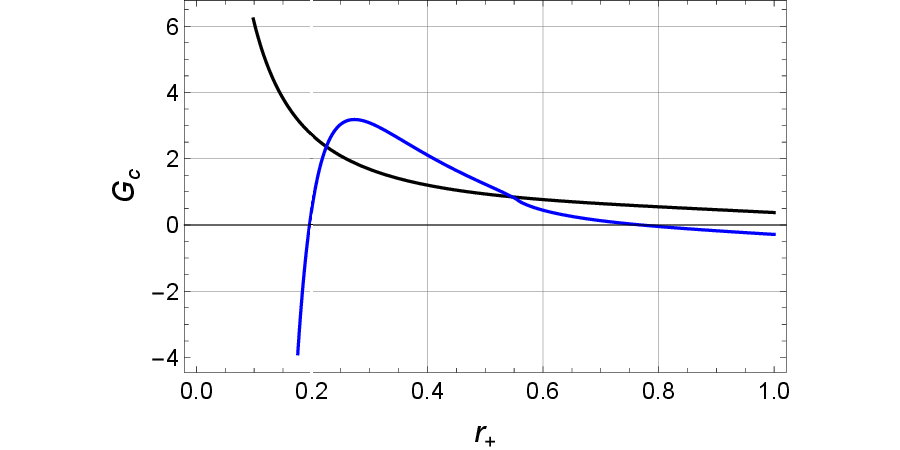}
    \end{tabular} 
    \caption{Logarithmic corrected Gibbs free energy vs event horizon radius of black holes with a CoS and PFDM. Here, $\alpha = 0$ is represented by black curve and $\alpha = 0.50$ is represented by blue curve. Left panel: $a = 0.75$ and $\lambda = 2$. Right panel: $a = 0.50$ and $\lambda = 1$. Both cases are plotted for constant values of $Q = l =1$.}
    \label{lncogpfdm}
\end{figure}

Logarithmic Corrected Gibbs free energy is given by:
\begin{equation}
 G_{c} = M_{c} - T_H S_{c},   
\end{equation}
where, $M_{c}$ is the corrected mass given in Eq. (\ref{M_cpfdm}) and $S_{c}$ is corrected entropy given in  Eq. (\ref{lncoentr}).
This simplifies to
\begin{eqnarray}
G_{c} &=&
\frac{r_{+}^{3}}{2l^{2}}
+\frac{Q^{2}+r_{+}^{2}(1-a)}{2r_{+}}
+\frac{\lambda}{2}\ln\!\left(\frac{r_{+}}{\lambda}\right)
-
\frac{r_{+}}{4}
\left(
1-a+\frac{3r_{+}^{2}}{l^{2}}
-\frac{Q^{2}}{r_{+}^{2}}
+\frac{\lambda}{r_{+}}
\right)
\nonumber\\
&+&
\alpha \Bigg[
-\frac{3r_{+}}{\pi l^{2}}
+\frac{4Q^{2}-3r_{+}\lambda}{12\pi r_{+}^{3}}
+
\frac{1}{4\pi r_{+}}
\left(
1-a+\frac{3r_{+}^{2}}{l^{2}}
-\frac{Q^{2}}{r_{+}^{2}}
+\frac{\lambda}{r_{+}}
\right)
\nonumber\\
&\times&
\ln\!\Bigg[
\frac{1}{16\pi}
\left(
1-a+\frac{3r_{+}^{2}}{l^{2}}
-\frac{Q^{2}}{r_{+}^{2}}
+\frac{\lambda}{r_{+}}
\right)^{2}
\Bigg]
\Bigg].
\label{G_cpfdm}
\end{eqnarray}
The behaviour of this corrected Gibbs free energy with respect to $r_+$ is depicted in
Fig. 
\ref{lncogpfdm}. We observe that the logarithmic corrected Gibbs free energy \(G_{c}\) exhibits significant variation with the event horizon radius \(r_{+}\) for black holes surrounded by a CoS and PFDM. The black curve represents the uncorrected case \((\alpha =0)\), while the blue curve corresponds to the logarithmically corrected case \((\alpha =0.50)\). In the left panel \((a=0.75,\ \lambda =2)\), the corrected Gibbs free energy initially increases from negative to positive values, attains a local maximum peak, and then decreases again toward the negative region as \(r_{+}\) increases. A similar behavior is observed in the right panel \((a=0.50,\ \lambda =1)\), although the variation becomes comparatively smoother and the deviation between the corrected and uncorrected curves decreases for larger horizon radii. The initial increase from negative to positive Gibbs free energy indicates that the black hole system becomes thermodynamically unstable and may undergo a phase transition to attain stability, whereas the subsequent decrease toward negative values suggests that the black hole configuration becomes thermodynamically preferred and globally stable, making the system less susceptible to further phase transitions in the large horizon radius regime.

From Figs.~\ref{lncohpfdm} and \ref{lncogpfdm}, we further observe that the behavior of the Gibbs free energy as a function of the event horizon radius for PFDM black holes with a CoS closely resembles the corresponding behavior of the Helmholtz free energy. In both cases, the thermodynamic potentials exhibit similar qualitative features, indicating a strong correlation in their thermodynamic evolution and phase structure with respect to the variation of the event horizon radius.
\subsection{Heat capacity and stability}

\begin{figure}[htbp]
\centering
\begin{tabular}{cc} 
    \includegraphics[width=0.52\textwidth]{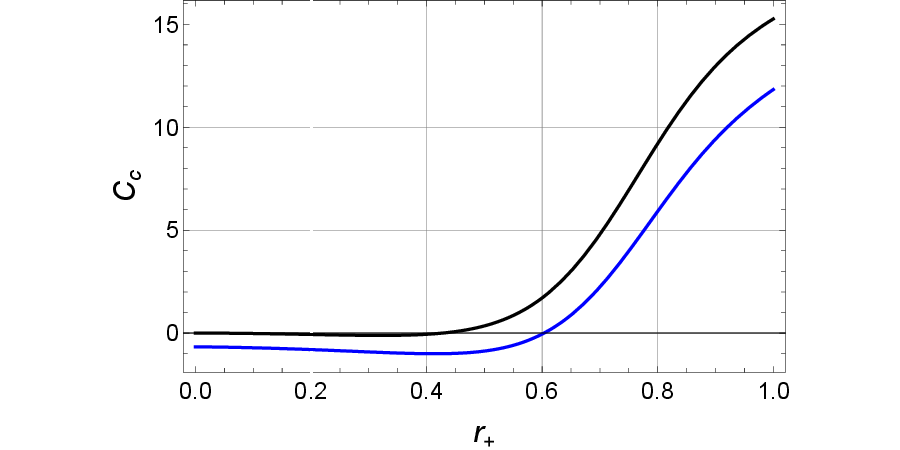}
    \includegraphics[width=0.52\textwidth]{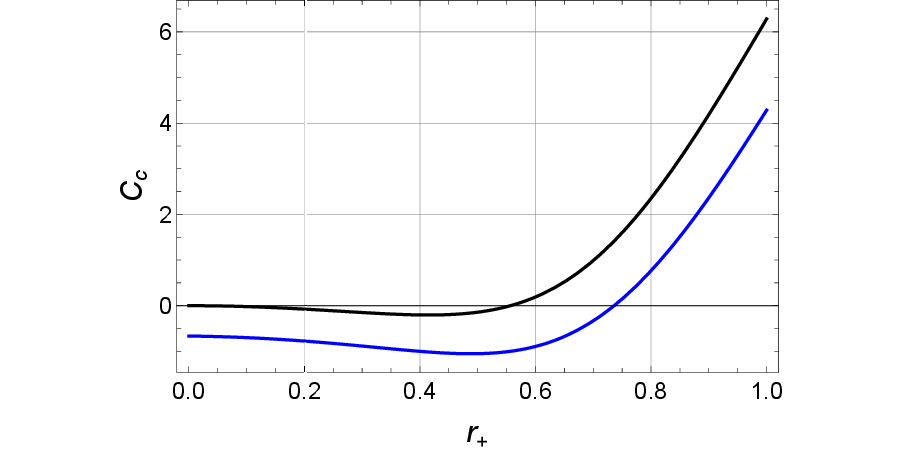}
    \end{tabular} 
    \caption{Logarithmic corrected heat capacity vs event horizon radius of black holes with a CoS and PFDM. Here, $\alpha = 0$ is represented by black curve and $\alpha = 0.50$ is represented by blue curve. Left panel: $a = 0.75$ and $\lambda = 2$. Right panel: $a = 0.50$ and $\lambda = 1$. Both cases are plotted for constant values of $Q = l =1$.}
    \label{lgcosppfdm}
\end{figure}

The logarithmic corrected heat capacity of PFDM black holes with a CoS is given by
\begin{equation}
 C_{c} = \frac{dM_{c}}{dT_{H}} = \frac{dM_{c}}{dr_{+}} \frac{dr_{+}}{dT_{H}}.  
\end{equation}
This leads to
\begin{eqnarray}
C_{c} &=&
\frac{
2\!\left[
3\pi r_{+}^{6}
+6\alpha r_{+}^{4}
+\pi l^{2}r_{+}^{2}
\left(
-Q^{2}+r_{+}^{2}(1-a)+\lambda r_{+}
\right)
+\alpha l^{2}(2Q^{2}-r_{+}\lambda)
\right]
}{
3r_{+}^{4}
+3Q^{2}l^{2}
+l^{2}r_{+}\!\left((a-1)r_{+}-2\lambda\right)
}.
\end{eqnarray}
To study the effects of perturbative correction on stability, we plot heat capacity with respect to horizon radius as depicted in Fig. \ref{lgcosppfdm}.   
Here, we observe that the logarithmic corrected specific heat capacity \(C_{c}\) exhibits significant dependence on the event horizon radius \(r_{+}\) for black holes surrounded by a CoS and PFDM. We note that the black curve represents the uncorrected case \((\alpha =0)\), while the blue curve corresponds to the logarithmically corrected case \((\alpha =0.50)\). In the left panel \((a=0.75,\ \lambda =2)\), we see that the corrected   heat capacity initially assumes negative values for small horizon radii, then increases rapidly, crosses the zero axis, and subsequently remains positive with a continuous increasing behavior as \(r_{+}\) increases. We observe a similar trend   in the right panel \((a=0.50,\ \lambda =1)\), although the growth of the corrected specific heat becomes comparatively smoother for larger horizon radii. The transition from negative to positive specific heat indicates that the black hole system evolves from a thermodynamically unstable phase to a stable phase. Furthermore, one can see that the positive and increasing nature of the corrected  heat capacity at larger values of \(r_{+}\) suggests that the PFDM black hole with a CoS acquires enhanced thermodynamic stability and becomes less susceptible to thermal fluctuations as the event horizon radius increases.   

\subsection{Perturbative corrected pressure and search for point of inflection of PFDM black holes with a CoS}
\begin{figure}[htbp]
\centering
\begin{tabular}{cc} 
    \includegraphics[width=0.52\textwidth]{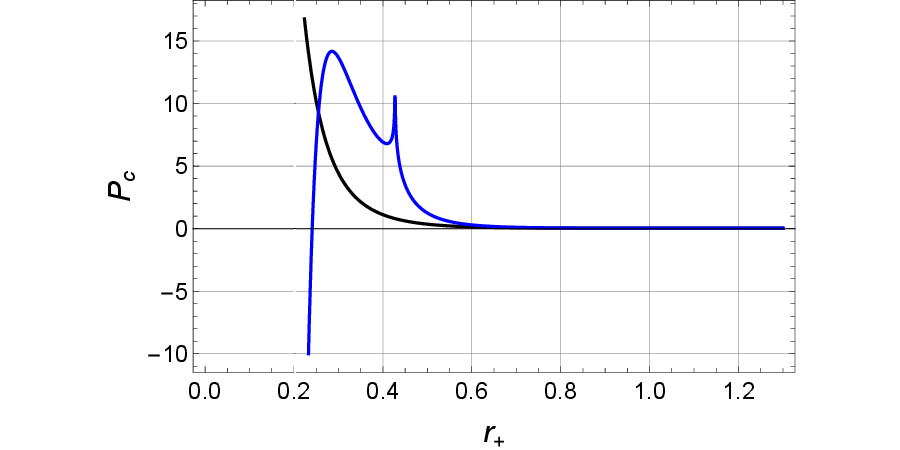}
    \includegraphics[width=0.52\textwidth]{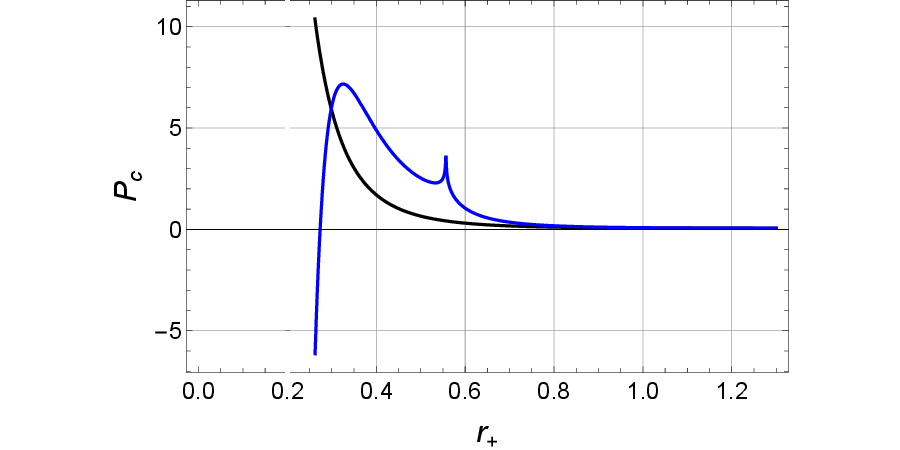} 
    \end{tabular} 
    \caption{Logarithmic corrected pressure vs event horizon radius of black holes with a CoS and PFDM. Here, $\alpha = 0$ is represented by black curve and $\alpha = 0.50$ is represented by blue curve. Left panel: $a = 0.75$ and $\lambda = 2$. Right panel: $a = 0.50$ and $\lambda = 1$. Both cases are plotted for constant values of $Q = l =1$.}
    \label{lncorpressure}
\end{figure}

The corrected pressure of the PFDM black holes with a CoS  is defined as
\begin{equation}
  P_c = -\left(\frac{\partial F_c}{\partial V}\right)_{T_{H}},
  \label{P_pfdm}
\end{equation}
where $F_{c}$ denotes the corrected Helmholtz free energy given by equation (\ref{F_cpfdm}), and volume $V = \frac{4}{3} \pi r^3_{+} $. 
Therefore,  using Eq. (\ref{F_cpfdm}) and Eq. (\ref{P_pfdm}),  the pressure of the PFDM black hole with a CoS can be computed  as 
\begin{eqnarray}
P_{c} &=&
\frac{
3r_{+}^{4}
+l^{2}\!\left[
3Q^{2}
+r_{+}\left((a-1)r_{+}-2\lambda\right)
\right]
}{
16\pi l^{2}r_{+}^{4}
}
\nonumber\\
&-& \alpha
\frac{
3r_{+}^{4}
+l^{2}\!\left[
3Q^{2}
+r_{+}\left((a-1)r_{+}-2\lambda\right)
\right]
}{
8\pi^{2}l^{2}r_{+}^{6}
}
\ln\!\left[\frac{1}{4\sqrt{\pi}}
\left(
-1+a
+\frac{Q^{2}}{r_{+}^{2}}
-\frac{3r_{+}^{2}}{l^{2}}
-\frac{\lambda}{r_{+}}
\right)
\right].
\label{P_c}
\end{eqnarray}
Since, $r_{+} = \sqrt[3]{\left(\frac{3V}{4\pi}\right)}$ therefore using this we get the expression of corrected pressure as 
\begin{eqnarray}
P_{c}
&=&
\frac{
9\,6^{\frac{1}{3}}V^{\frac{4}{3}}
+
24\pi^{\frac{4}{3}}l^{2}Q^{2}
+
2( -1+a)(6\pi)^{\frac{2}{3}}l^{2}V^{\frac{2}{3}}
-
8\,6^{\frac{1}{3}}\pi l^{2}V^{\frac{1}{3}}\lambda
}{
288\,l^{2}\pi^{\frac{4}{3}}V^{2}
}
\nonumber\\
&\times&
\Bigg[
6^{\frac{2}{3}}\pi^{\frac{1}{3}}V^{\frac{2}{3}}
+
4\alpha \ln(16\pi)
-
8\alpha
\ln\!\Bigg(
1-a
-\frac{2\,6^{\frac{1}{3}}\pi^{\frac{2}{3}}Q^{2}}{3V^{\frac{2}{3}}}
+
\frac{
\left(\frac{6}{\pi}\right)^{\frac{2}{3}}
(9V+4l^{2}\pi\lambda)
}{
12l^{2}V^{\frac{1}{3}}
}
\Bigg)
\Bigg].
\label{P_c(V)}
\end{eqnarray}
From the figure \ref{lncorpressure}, we observe  that the logarithmic corrected pressure \(P_{c}\) shows significant dependence on the event horizon radius \(r_{+}\) for black holes surrounded by a CoS and PFDM. The black curve corresponds to the uncorrected case \((\alpha =0)\), whereas the blue curve represents the logarithmically corrected case \((\alpha =0.50)\). In the left panel \((a=0.75,\ \lambda =2)\), we find that the corrected pressure initially increases and attains a local maximum, then decreases and subsequently develops a pronounced positive peak before decreasing again and gradually coinciding with the uncorrected pressure curve for larger values of \(r_{+}\). In the right panel \((a=0.50,\ \lambda =1)\), one can observe that a similar qualitative behavior is observed, although the magnitude of the positive peak becomes comparatively smaller. The comparatively larger peak is obtained for the CoS parameter \(a=0.75\) and scale parameter \(\lambda =2\) indicates that these parameters strongly influence the thermodynamic pressure in the presence of thermal fluctuation corrections. Furthermore, the convergence of the corrected and uncorrected pressure curves at large horizon radii suggests that the contribution of logarithmic corrections becomes negligible in the large black hole regime. 

\begin{figure}[htbp]
\centering
\begin{tabular}{cc} 
    \includegraphics[width=0.52\textwidth]{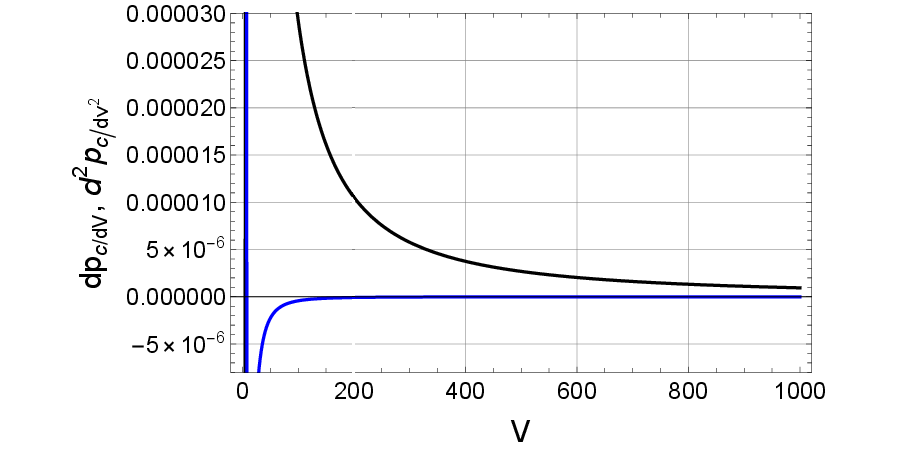}
    \includegraphics[width=0.52\textwidth]{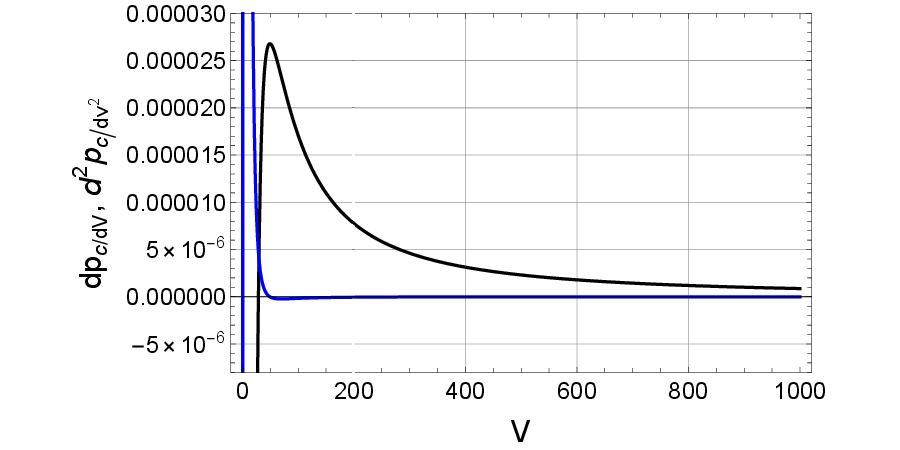} 
    \end{tabular} 
    \caption{First and second order derivatives of logarithmic corrected pressure vs volume of black holes with a CoS and PFDM. Here, $\left(\frac{\partial P}{\partial V} \right)_{T_H = T_c} = 0$ is represented by black curve and $\left(\frac{\partial^2 P}{\partial V^2} \right)_{T_H = T_c} = 0$ is represented by blue curve. Left panel: $a = 0.75$ and $\lambda = 2$. Right panel: $a = 0.50$ and $\lambda = 1$.  Both cases are plotted for constant values of $Q = l =1$.}
    \label{lndericorpressure}
\end{figure}

From the Fig. \ref{lndericorpressure}, we observe that the first and second order derivatives of the logarithmic corrected pressure with respect to the thermodynamic volume \(V\) exhibit distinct behaviors for black holes surrounded by a CoS and PFDM. The black curve represents the condition \(\left(\frac{\partial P}{\partial V}\right)_{T_H=T_c}=0\), while the blue curve corresponds to \(\left(\frac{\partial^2 P}{\partial V^2}\right)_{T_H=T_c}=0\). In the left panel \((a=0.75,\ \lambda =2)\), both curves vary continuously with the volume; however, they do not intersect each other at the common value of zero within the considered range of \(V\). A similar behavior is observed in the right panel \((a=0.50,\ \lambda =1)\), where the two curves again remain separated throughout the plotted region. The absence of a common intersection point between the first and second derivative curves indicates that the simultaneous conditions \(\left(\frac{\partial P}{\partial V}\right)_{T_H=T_c}=0\) and \(\left(\frac{\partial^2 P}{\partial V^2}\right)_{T_H=T_c}=0\) are not satisfied for any critical volume \(V_{c}\). Therefore, no critical point exists for the logarithmically corrected pressure within the considered parameter range, suggesting the absence of standard Van der Waals type critical behavior in the PFDM black hole system with a CoS under thermal fluctuation corrections.

\section{Non-perturbative corrections to thermodynamics of PFDM black holes with a CoS}\label{sec4}
In this section, we discuss the effects of non-perturbative correction to the entropy  on the thermal properties of black holes with a CoS and PFDM. 
The general expression for non-perturbative (exponential) correction to entropy is given by \cite{40}  
\begin{equation}
 S_{\mathrm{exp}} = \pi r_{+}^2 + \eta \, e^{-\pi r_{+}^2} ,
 \label{Exp_entr}
\end{equation}
where $\eta$ is a correction factor. 

\begin{figure}[htbp]
    \centering
    \includegraphics[width=0.5\linewidth]{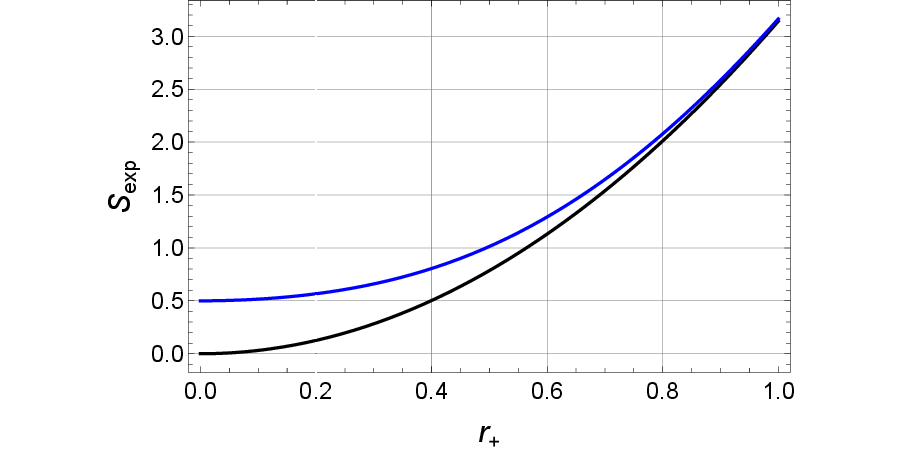}
    \caption{Non-perturbative corrected entropy vs event horizon radius of black holes with a CoS and PFDM. Here, $\eta = 0$ is represented by black curve and $\eta = 0.5$ is represented by blue curve.}
    \label{expcoentr}
\end{figure}

The effects of non-perturbative correction on entropy is depicted in  the figure \ref{expcoentr}.  Here, we see that the non-perturbative corrected entropy $S_{\mathrm{exo}}$ increases monotonically with the event horizon radius $r_{+}$ for both values of the parameter $\eta$. The black curve, corresponding to $\eta = 0$, and the blue curve, corresponding to $\eta = 0.5$, show that the presence of PFDM enhances the entropy, particularly in the small horizon radius region where the deviation between the two curves is more pronounced. As $r_{+}$ increases, the two curves gradually approach each other, indicating that the effect of $\eta$ becomes weaker for larger black holes. This behavior suggests that the non-perturbative corrections significantly modify the thermodynamic properties of small black holes, while their influence diminishes in the large horizon radius limit.

\subsection{Exponentially corrected mass}
\begin{figure}[htbp]
\centering
\begin{tabular}{cc} 
    \includegraphics[width=0.52\textwidth]{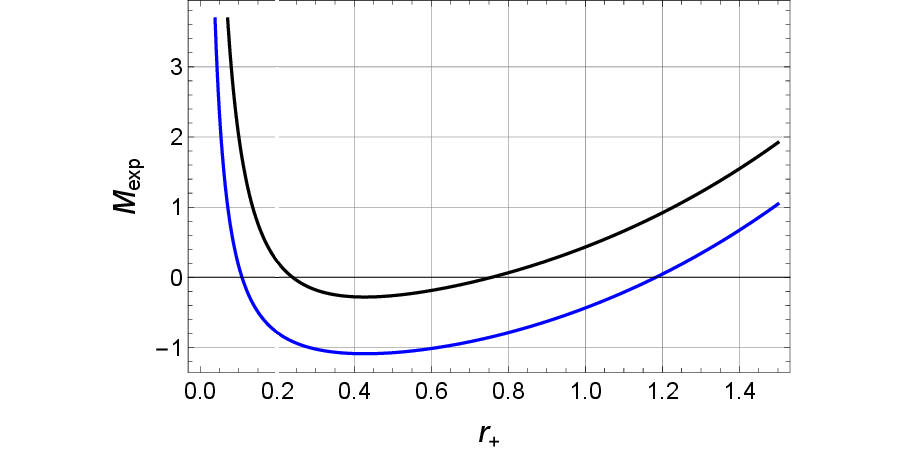}
    \includegraphics[width=0.52\textwidth]{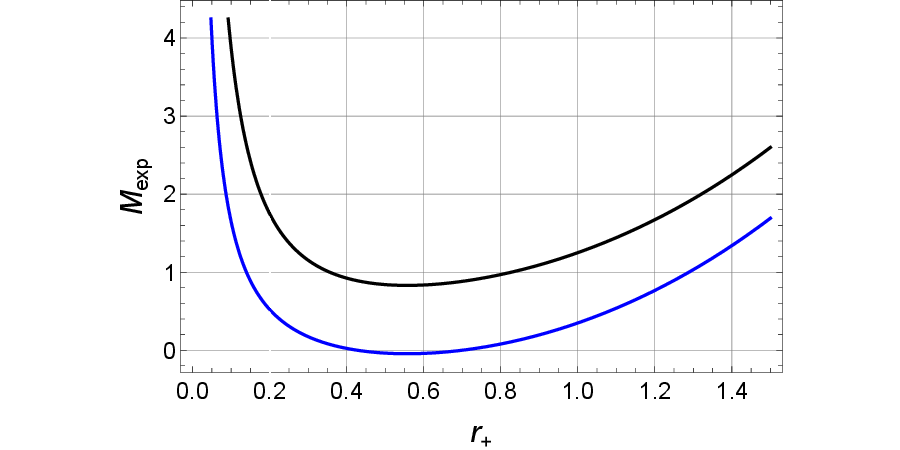}
    \end{tabular} 
    \caption{Exponentially corrected mass vs event horizon radius of black holes with a CoS and PFDM. Here, $\eta = 0$ is represented by black curve and $\eta = 0.50$ is represented by blue curve. Left panel: $a = 0.75$ and $\lambda = 2$. Right panel: $a = 0.50$ and $\lambda = 1$. The parameters are fixed in both cases as $Q = l =1$.}
    \label{M_expc}
\end{figure}

The non-perturbative corrected mass can be calculated from the following standard definition:
\begin{equation}
 M_{\mathrm{exp}} = \int T_{H} \, dS_{\mathrm{exp}}. 
\end{equation}
Exploiting the value of $T_{H}$ from (\ref{T_pfdmHaw}) and $S_{\text{exc}}$ from (\ref{Exp_entr}), this expression leads to
\begin{eqnarray}
M_{\mathrm{exp}}
&=&
\frac{e^{-\pi r_{+}^{2}}}{8\pi l^{2}r_{+}}
\Bigg[
e^{\pi r_{+}^{2}}
\left\{
4\pi\left(r_{+}^{4}+l^{2}\left(Q^{2}+(a-1)r_{+}^{2}\right)\right)
\right.\nonumber\\
&&
+\left.\,\eta
\left(
6r_{+}^{2}
-4\pi l^{2}Q^{2}
-\left[3+2\pi l^{2}\left(1-a+2\pi Q^{2}\right)\right]
e^{\pi r_{+}^{2}}r_{+}
\operatorname{Erf}\!\left(\sqrt{\pi}\,r_{+}\right)\right.\right.
\nonumber\\
&& \left. \left.
-2\pi l^{2}r_{+}\lambda\,
e^{\pi r_{+}^{2}}
\operatorname{Ei}\!\left(-\pi r_{+}^{2}\right)
\right)
+\,4\pi l^{2}r_{+}\lambda\,
e^{\pi r_{+}^{2}}
\ln\!\left(\frac{r_{+}}{\lambda}\right)
\right\}
\Bigg].
\label{M_expcpfdm}
\end{eqnarray}
 The effects of non-perturbative correction on mass of black hole is depicted in  the figure \ref{M_expc}. 
We observe that the exponentially corrected mass $M_{\mathrm{exp}}$ increases with the event horizon radius $r_{+}$ for both values of the dark matter parameter $\eta$, indicating that larger black holes possess greater thermodynamic mass. The black curve corresponds to $\eta = 0$, while the blue curve represents $\eta = 0.50$. In both panels, the inclusion of PFDM enhances the black hole mass, particularly in the small horizon radius region where the deviation between the two curves is more significant. In the left panel, corresponding to $a = 0.75$ and $\lambda = 2$, the mass grows more rapidly with $r_{+}$ compared to the right panel with $a = 0.50$ and $\lambda = 1$, showing that the parameters $a$ and $\lambda$ strongly influence the thermodynamic behavior of the system. As the horizon radius increases, the difference between the two curves gradually decreases, implying that the effect of the dark matter parameter becomes less dominant for large black holes. The overall positive behavior of $M_{\mathrm{exp}}$ suggests that the exponentially corrected black hole configuration remains thermodynamically well behaved in the considered parameter range.

\subsection{Exponentially corrected Helmholtz free energy}
\begin{figure}[htbp]
\centering
\begin{tabular}{cc} 
    \includegraphics[width=0.52\textwidth]{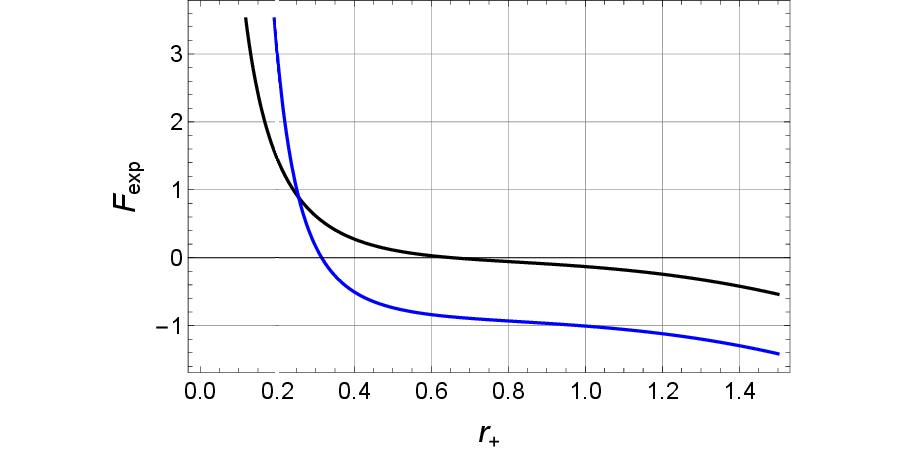}
    \includegraphics[width=0.52\textwidth]{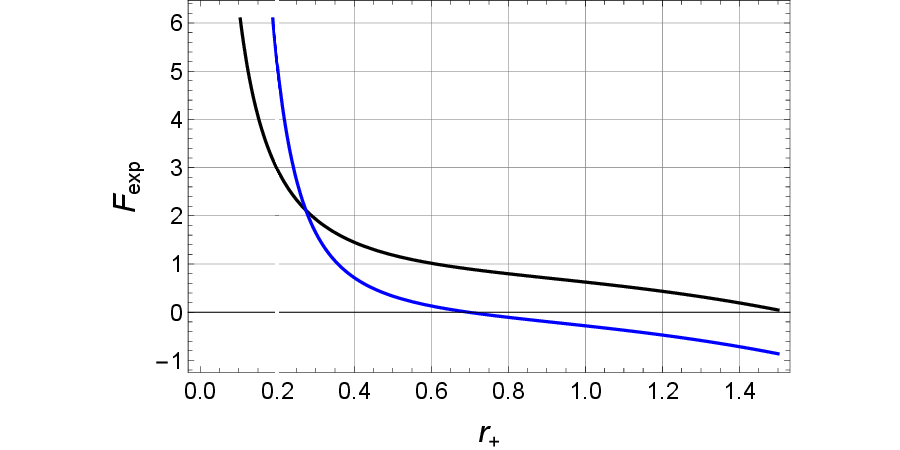}
    \end{tabular} 
    \caption{Exponentially corrected Helmholtz free energy vs event horizon radius of black holes with a CoS and PFDM. Here, $\eta = 0$ is represented by black curve and $\eta = 0.50$ is represented by blue curve. Left panel: $a = 0.75$ and $\lambda = 2$. Right panel: $a = 0.50$ and $\lambda = 1$. The parameters are fixed in both cases as $Q = l =1$.}
    \label{F_expc}
\end{figure}

To compute the corrected Helmholtz free energy, we utilize the following expression:  
\begin{equation}
 F_{\mathrm{exp}} = - \int S_{\mathrm{exp}} \, dT_{H}.
 \end{equation}
This results to
\begin{eqnarray}
F_{\mathrm{exp}}
&=&
-\frac{1}{8\pi r_{+}^{3}}
\Bigg[
2\pi r_{+}^{2}
\left(
-3Q^{2}
+(a-1)r_{+}^{2}
+\frac{r_{+}^{4}}{l^{2}}
-2r_{+}\lambda
\ln\!\frac{r_{+}}{\lambda}
\right)
\nonumber\\
&&
+\eta
\Bigg\{ +\,2\pi r_{+}^{3}\lambda\,
\operatorname{Ei}\!\left(-\pi r_{+}^{2}\right) 
+\frac{r_{+}^{3}}{l^{2}}
\left[
3+2\pi l^{2}(1-a+2\pi Q^{2})
\right]
\operatorname{Erf}\!\left(\sqrt{\pi}\,r_{+}\right)
\nonumber\\
&&
+2e^{-\pi r_{+}^{2}}
\left[
Q^{2}\!\left(-1+2\pi r_{+}^{2}\right)
+r_{+}^{2}(1-a)
+r_{+}\lambda
\right]
\Bigg\}
\Bigg].
\label{F_expcpfdm}
\end{eqnarray}
To see the effect of non-perturbative correction on Helmholtz free energy, we plot  the graph \ref{F_expc}. From the plot, we observe that the exponentially corrected Helmholtz free energy $F_{\mathrm{exp}}$ exhibits a non-linear dependence on the event horizon radius $r_{+}$ for both values of the dark matter parameter $\eta$.   In both panels, the free energy initially decreases in the small horizon radius region and then gradually increases as $r_{+}$ becomes larger, indicating a change in the thermodynamic behavior of the black hole system. The inclusion of PFDM increases the magnitude of the Helmholtz free energy, particularly for smaller black holes, where the separation between the two curves is more prominent. Comparing the two panels, we find that the left panel with $a = 0.75$ and $\lambda = 2$ shows a stronger variation in free energy than the right panel with $a = 0.50$ and $\lambda = 1$, demonstrating that the CoS parameter $a$ and the parameter $\lambda$ significantly affect the thermodynamic structure of the black hole. For larger values of $r_{+}$, the curves tend to approach each other, implying that the effect of the dark matter parameter becomes less significant in the large black hole limit.
 \subsection{Exponentially corrected Gibbs free energy}
\begin{figure}[htbp]
\centering
\begin{tabular}{cc} 
    \includegraphics[width=0.52\textwidth]{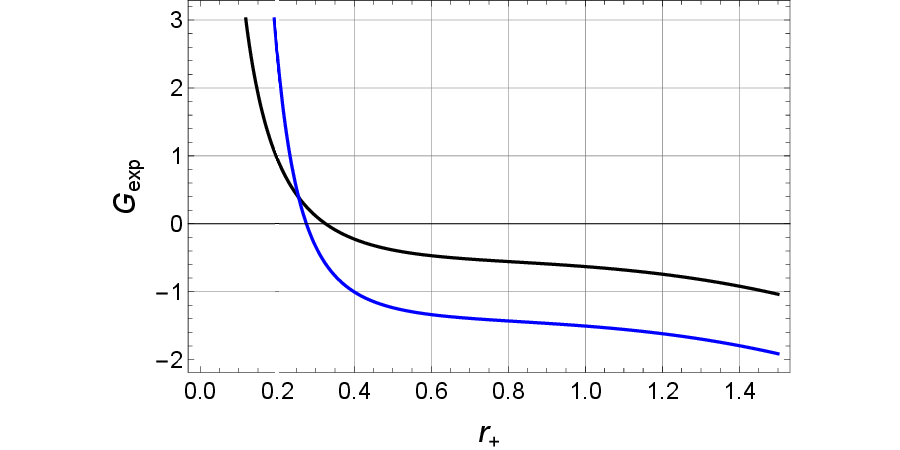}
    \includegraphics[width=0.52\textwidth]{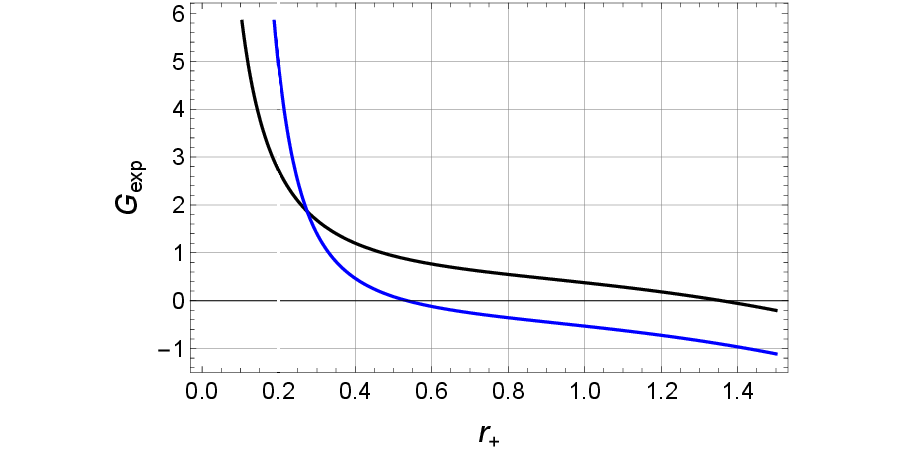}
    \end{tabular} 
    \caption{Exponentially corrected Gibbs free energy vs event horizon radius of black holes with a CoS and PFDM. Here, $\eta = 0$ is represented by black curve and $\eta = 0.50$ is represented by blue curve. Left panel: $a = 0.75$ and $\lambda = 2$. Right panel: $a = 0.50$ and $\lambda = 1$. The parameters are fixed in both cases as $Q = l =1$.}
    \label{G_expc}
\end{figure}

The non-perturbative exponentially corrected Gibbs free energy can be estimated from 
the standard definition,
\begin{equation}
 G_{\mathrm{exp}} = M_{\mathrm{exp}} - T_{H} S_{\mathrm{exp}}.   
\end{equation}
The explicit expressions for $M_{\mathrm{exp}}$ and $S_{\mathrm{exp}}$  are given in Eq. (\ref{M_expcpfdm}) and  Eq. (\ref{Exp_entr}), respectively.
Plugging these values of  $M_{\mathrm{exp}}$,  $T_{H}$ and $S_{\mathrm{exp}}$, the above relation leads to the exponentially corrected Gibbs free energy  as follows
\begin{eqnarray}
G_{\mathrm{exp}}
&=&
\frac{1}{2l^{2}}
\left(
r_{+}^{3}
+l^{2}(a-1)r_{+}
+\frac{l^{2}Q^{2}}{r_{+}}
+l^{2}\lambda \ln\!\frac{r_{+}}{\lambda}
\right)
-\frac{1}{4r_{+}}
\left(
1-a+\frac{3r_{+}^{2}}{l^{2}}
-\frac{Q^{2}}{r_{+}^{2}}
+\frac{\lambda}{r_{+}}
\right)
r_{+}^{2}
\nonumber\\
&
+&\eta\,e^{-\pi r_{+}^{2}}
\left[
\frac{1}{8\pi l^{2}r_{+}}
\left(
6r_{+}^{2}
-4\pi l^{2}Q^{2}
-\Bigl[3+2\pi l^{2}(1-a+2\pi Q^{2})\Bigr]
e^{\pi r_{+}^{2}}r_{+}
\operatorname{Erf}\!\left(\sqrt{\pi}\,r_{+}\right)
\right.\right. \nonumber\\
&
-&\left.\left. 2\pi l^{2}r_{+}\lambda\,
e^{\pi r_{+}^{2}}
\operatorname{Ei}\!\left(-\pi r_{+}^{2}\right)
\right)
-\frac{1}{4\pi r_{+}}
\left(
1-a+\frac{3r_{+}^{2}}{l^{2}}
-\frac{Q^{2}}{r_{+}^{2}}
+\frac{\lambda}{r_{+}}
\right)
\right].
\label{G_expcpfdm}
\end{eqnarray}
The variation of exponentially corrected Gibbs free energy with respect to horizon radius is depicted in Fig.  \ref{G_expc}. Here, we observe that the exponentially corrected Gibbs free energy $G_{\mathrm{exp}}$ shows a non-trivial dependence on the event horizon radius $r_{+}$ for both values of the dark matter parameter $\eta$.  In both panels, the Gibbs free energy decreases in the small horizon radius region and then gradually increases as the black hole size grows, indicating the presence of distinct thermodynamic phases. The inclusion of PFDM modifies the Gibbs free energy significantly for smaller black holes, where the separation between the two curves becomes more pronounced. In the left panel, corresponding to $a = 0.75$ and $\lambda = 2$, the variation of $G_{\mathrm{exp}}$ is stronger than in the right panel with $a = 0.50$ and $\lambda = 1$, showing that the CoS parameter $a$ and the parameter $\lambda$ strongly influence the thermodynamic behavior of the system. For larger values of $r_{+}$, the curves tend to merge, implying that the contribution of the dark matter parameter becomes weaker in the large black hole limit. The smooth behavior of the Gibbs free energy further suggests that the exponentially corrected black hole system remains thermodynamically stable within the considered parameter range.

\subsection{Exponentially corrected heat capacity and stability}
\begin{figure}[htbp]
\centering
\begin{tabular}{cc} 
    \includegraphics[width=0.52\textwidth]{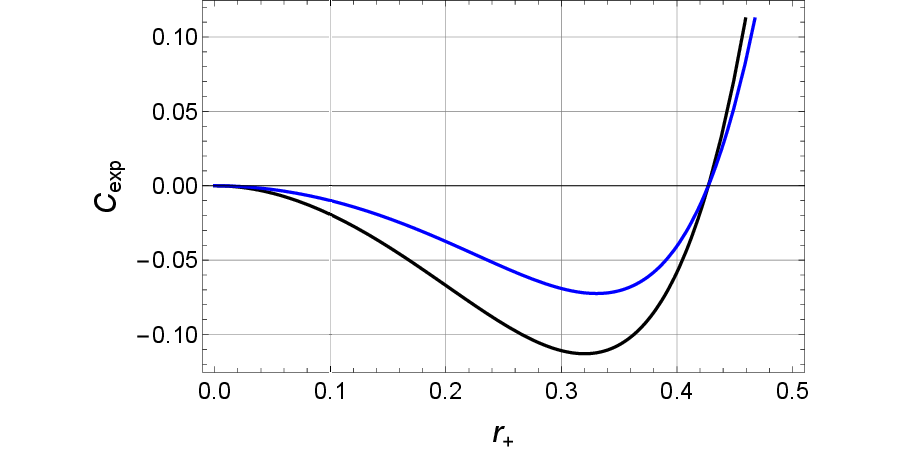}
    \includegraphics[width=0.52\textwidth]{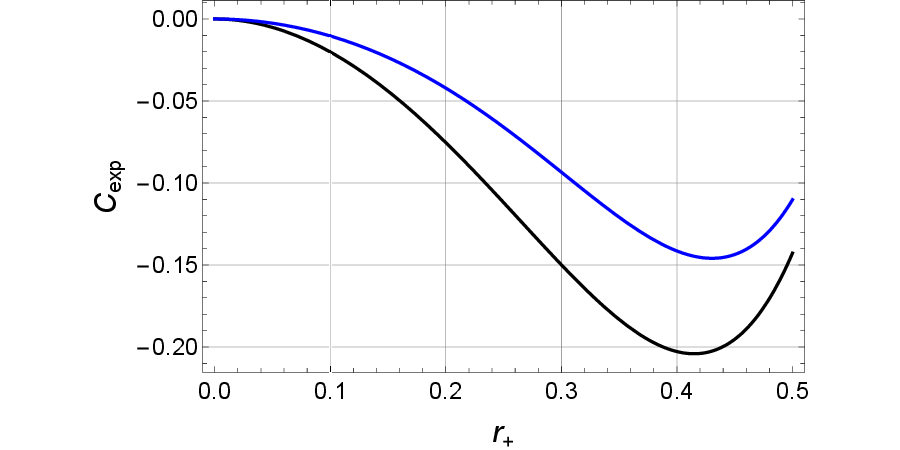}
    \end{tabular} 
    \caption{Exponentially corrected heat capacity vs event horizon radius of black holes with a CoS and PFDM. Here, $\eta = 0$ is represented by black curve and $\eta = 0.50$ is represented by blue curve. Left panel: $a = 0.75$ and $\lambda = 2$. Right panel: $a = 0.50$ and $\lambda = 1$. The parameters are fixed in both cases as $Q = l =1$.}
    \label{Exp_spec}
\end{figure}

The exponentially corrected heat capacity after introducing non-perturbative correction  can be calculated from the following relation: 
\begin{equation}
 C_{\mathrm{exp}} = \frac{dM_{\mathrm{exp}}}{dT_{H}} = \frac{dM_{\mathrm{exp}}}{dr_{+}} \frac{dr_{+}}{dT_{H}}.  
\end{equation}
This leads to
\begin{equation}
C_{\mathrm{exp}}
=
\frac{
2\pi r_{+}^{2}
\,e^{-\pi r_{+}^{2}}
\left(e^{\pi r_{+}^{2}}-\eta\right)
\left[
3r_{+}^{4}
+l^{2}\left(-Q^{2}+(1-a)r_{+}^{2}+r_{+}\lambda\right)
\right]
}{
3r_{+}^{4}
+3Q^{2}l^{2}
+l^{2}r_{+}\left((a-1)r_{+}-2\lambda\right)
}.
\end{equation}
Here, from the left panel of figure \ref{Exp_spec}, we see that heat capacity first decreases to negative value and then increases to positive value thus suggesting that there is a phase transition phenomenon occurring in our black hole system and black hole goes from less stable to a more stable phase. Whereas, from the right panel we observe that heat capacity shows behaviour of increasingly negative value for the small range of event horizon radius.
We observe that the exponentially corrected heat capacity $C_{\mathrm{exp}}$ exhibits a highly sensitive dependence on the event horizon radius $r_{+}$ for both values of the dark matter parameter $\eta$.  In both panels, the heat capacity changes significantly with increasing horizon radius, indicating the presence of different thermodynamic stability regions. The positive values of $C_{\mathrm{exp}}$ correspond to thermodynamically stable black hole configurations, whereas negative values indicate unstable phases. The divergence behavior of the heat capacity suggests the existence of second-order phase transitions in the black hole system. The inclusion of PFDM modifies the location and magnitude of these transition points, particularly in the small horizon radius region where the separation between the black and blue curves becomes more pronounced. Comparing the two panels, the left panel with $a = 0.75$ and $\lambda = 2$ shows stronger variations and sharper transitions than the right panel with $a = 0.50$ and $\lambda = 1$, demonstrating that the CoS parameter $a$ and the parameter $\lambda$ significantly affect the thermodynamic stability of the black hole. For larger values of $r_{+}$, the curves gradually approach each other, indicating that the influence of the dark matter parameter becomes weaker in the large black hole regime.
\subsection{Non-perturbative corrected pressure and search for point of inflection of PFDM black holes with a CoS}
We calculate the exponentially corrected pressure of PFDM black holes in the presence of a  CoS  from the derivative of the exponentially corrected Helmholtz free energy  with respect to the thermodynamic volume    as
\begin{equation}
  P_{\mathrm{exp}} = -\left(\frac{\partial F_{\mathrm{exp}}}{\partial V}\right)_{T_{H}},
  \label{P_expfdm}
\end{equation}
where  volume has the following expression: $V = \frac{4}{3} \pi r^3_{+} $.

 Utilizing Eqs. (\ref{F_expcpfdm}) and  (\ref{P_expfdm}),  the pressure of the PFDM black hole with a CoS is computed by  
\begin{equation}
P_{\mathrm{exp}}
=
\frac{e^{-\pi r_{+}^{2}} \left( e^{\pi r_{+}^{2}} \pi r_{+}^{2} + \eta \right)
\left[ 3 r_{+}^{4} + l^{2} \left( 3 Q^{2} + r_{+} \left( (a-1) r_{+} - 2 \lambda \right) \right) \right]}
{16\, l^{2}\, \pi^{2}\, r_{+}^{6}}.
\end{equation}
Since  $r_{+} = \sqrt[3]{\left(\frac{3V}{4\pi}\right)}$,  the expression of corrected pressure can be expressed in terms of volume as
\begin{equation}
\begin{aligned}
P_{\mathrm{exp}}
&=
\frac{
e^{-\frac{1}{2} 3^{2/3} \left( \frac{\pi}{2} \right)^{1/3} V^{2/3}}
}{288\, l^{2}\, \pi^{4/3}\, V^{2}} 
\left[
6^{2/3}
e^{\frac{1}{2} 3^{2/3} \left( \frac{\pi}{2} \right)^{1/3} V^{2/3}}
\pi^{1/3} V^{2/3}
+ 4 \eta
\right]
\\[0.6em]
&\quad \times
\left[
9\,6^{1/3} V^{4/3}
+ 2 l^{2}
\left(
12 \pi^{4/3} Q^{2}
+ (a-1) (6\pi)^{2/3} V^{2/3}
- 4\,6^{1/3} \pi \lambda V^{1/3}
\right)
\right].
\end{aligned}
\label{P_expc}
\end{equation}
 Now, to do comparative analysis, we plot   figure \ref{exdericorpressure}.
\begin{figure}[htbp]
\centering
\begin{tabular}{cc} 
    \includegraphics[width=0.52\textwidth]{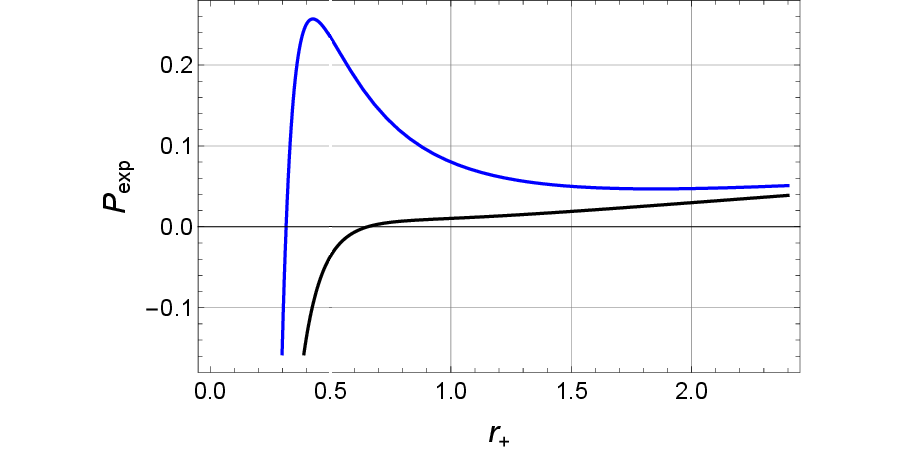}
    \includegraphics[width=0.52\textwidth]{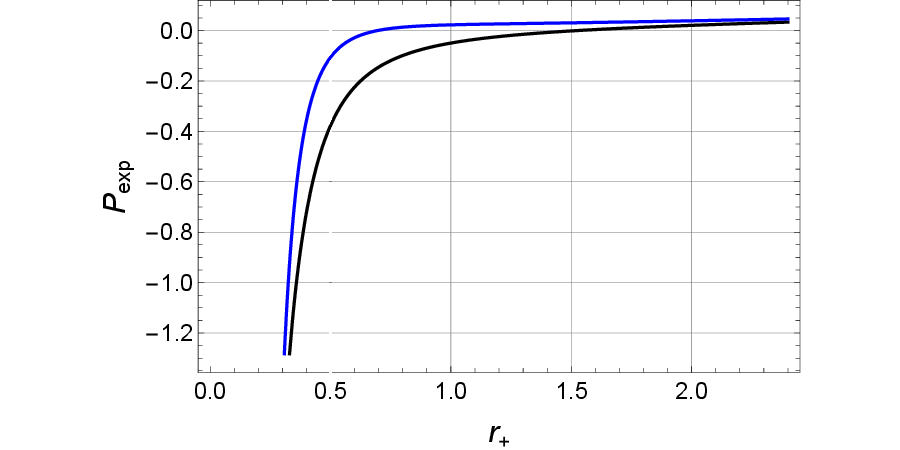} 
    \end{tabular} 
    \caption{Exponentially corrected pressure vs event horizon radius of black holes with a CoS and PFDM. Here, $\alpha = 0$ is represented by black curve and $\alpha = 0.50$ is represented by blue curve. Left panel: $a = 0.75$ and $\lambda = 2$. Right panel: $a = 0.50$ and $\lambda = 1$. The parameters are fixed in both cases as $Q = l =1$.}
    \label{excopressure}
\end{figure}
We observe that the exponentially corrected pressure  increases with the event horizon radius   for both values of the correction parameter $\alpha$.  In both panels, the presence of the exponential correction enhances the pressure, particularly in the small horizon radius region where the deviation between the two curves is more significant. As the event horizon radius increases, the pressure grows smoothly and the difference between the curves gradually decreases, indicating that the effect of the correction parameter becomes weaker for larger black holes. Comparing the two panels, the left panel with $a = 0.75$ and $\lambda = 2$ exhibits a stronger variation in pressure than the right panel with $a = 0.50$ and $\lambda = 1$, showing that the CoS parameter $a$ and the parameter $\lambda$ significantly influence the thermodynamic behavior of the black hole system. The overall positive behavior of the pressure suggests that the exponentially corrected PFDM black holes with a CoS remain thermodynamically consistent within the considered parameter range.
\begin{figure}[htb]
\centering
\begin{tabular}{cc} 
    \includegraphics[width=0.52\textwidth]{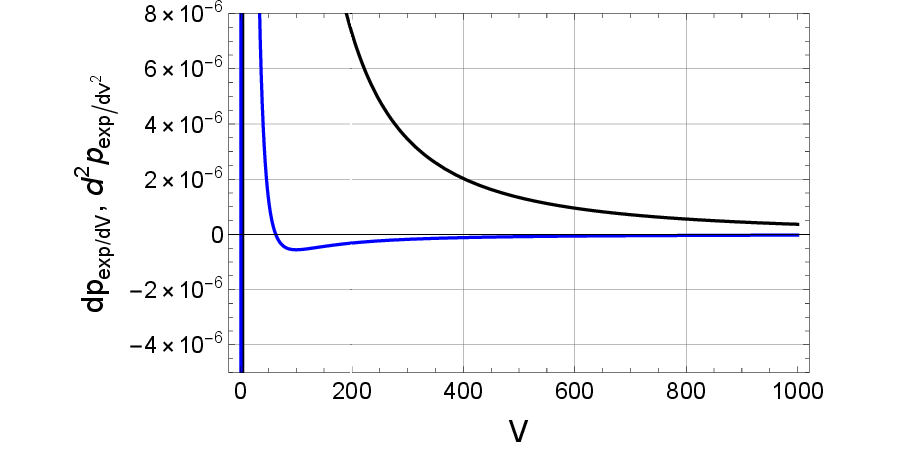}
    \includegraphics[width=0.52\textwidth]{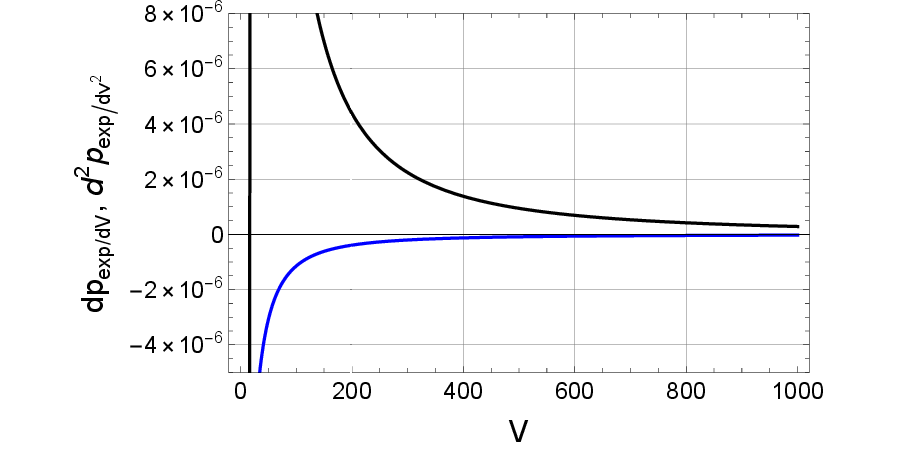} 
    \end{tabular} 
    \caption{First and second order derivatives of exponentially corrected pressure vs volume of black holes with a CoS and PFDM. Here, $\left(\frac{\partial P_{\text{exc}}}{\partial V} \right)_{T_H = T_c} = 0$ is represented by black curve and $\left(\frac{\partial^2 P_{\text{exc}}}{\partial V^2} \right)_{T_H = T_c} = 0$ is represented by blue curve. Left panel: $a = 0.75$ and $\lambda = 2$. Right panel: $a = 0.50$ and $\lambda = 1$. The parameters are fixed in both cases as $Q = l =1$.}
    \label{exdericorpressure}
\end{figure}

Now, employing Eq. (\ref{P_expc}), we compute the first and second order derivatives of corrected pressure with respect to the thermodynamic volume and plot the figure \ref{exdericorpressure}. Here, it is evident that curves defined by $\left(\frac{\partial P}{\partial V} \right)_{T_H = T_c} = 0$ and $\left(\frac{\partial P}{\partial V} \right)_{T_H = T_c} = 0$ do not intersect for any common value of $V$ within the domain under consideration. This observation indicates that the inflection point conditions cannot be satisfied simultaneously. Therefore, the system does not admit a critical point, and no critical volume $V_{c}$ can be found in this case.

\section{Concluding remarks} \label{sec5} 

In this work, we have computed the perturbative and non-perturbative corrected thermodynamic properties of charged AdS black holes surrounded by both CoS  and  PFDM. Starting from the Einstein field equations in the presence of electromagnetic fields, PFDM, and CoS contributions, we revisited the corresponding black hole geometry and investigated its thermodynamic behavior under statistical entropy corrections. 

We have first analyzed the perturbative logarithmic corrections to the Bekenstein--Hawking entropy arising due to thermal fluctuations around equilibrium. Using the corrected entropy, we have derived explicit analytical expressions for the corrected mass, Helmholtz free energy, Gibbs free energy, heat capacity, and thermodynamic pressure. Our analysis shows that logarithmic corrections significantly modify the thermodynamic behavior in the small black hole regime. In particular, the corrected entropy exhibits divergence behavior for small horizon radius, while the corrected heat capacity changes from negative to positive values, indicating a transition from thermodynamically unstable to stable phases. Furthermore, the Gibbs and Helmholtz free energies reveal possible phase transition structures and enhanced global stability for larger black holes. However, by investigating  the first and second derivatives of pressure with respect to thermodynamic volume, we have found the absence of a common inflection point, suggesting that the logarithmically corrected PFDM black holes with a CoS do not exhibit standard Van der Waals type criticality within the considered parameter range. 

Subsequently, we have explored the effects of non-perturbative exponential corrections to entropy. We have obtained the corresponding exponentially corrected thermodynamic quantities and examined their physical behavior graphically. The results indicated that exponential corrections predominantly affect the small horizon radius region, whereas their influence becomes negligible for large black holes. The corrected mass and pressure remain positive throughout the considered domain, indicating thermodynamic consistency of the system. Moreover, the behavior of the exponentially corrected heat capacity suggests the existence of second-order phase transitions and demonstrates that the CoS parameter and PFDM scale parameter strongly influence the thermodynamic stability of the black hole configuration. Similar to the perturbative case, the inflection point analysis confirms the absence of critical behavior associated with Van der Waals type phase transitions. 

Overall, we have demonstrated that  both perturbative and non-perturbative entropy corrections play an important role in modifying the thermal properties and stability structure of PFDM black holes surrounded by a CoS, especially in the quantum regime of small horizon radius. The combined effects of the CoS parameter, dark matter scale parameter, electric charge, and thermal fluctuation corrections enrich the phase structure and thermodynamic evolution of the system. These results may provide useful insights into quantum corrected black hole thermodynamics in dark matter dominated backgrounds and could be extended further in the context of modified gravity, holography, or quasinormal mode analysis in future studies.

\end{document}